# Observing and modeling the sequential pairwise reactions that drive solid-state ceramic synthesis

Akira Miura,[1*†] Christopher J. Bartel,[2†] Yusuke Goto,[3] Yoshikazu Mizuguchi,[3] Chikako Moriyoshi,[4] Yoshihiro Kuroiwa,[4] Yongming Wang,[5] Toshie Yaguchi,[6] Manabu Shirai,[6] Masanori Nagao,[7] Nataly Carolina Rosero-Navarro,[1] Kiyoharu Tadanaga,[1] Gerbrand Ceder,[2,8] Wenhao Sun[9*]

[*]Correspondence to: amiura@eng.hokudai.ac.jp (A.M.), whsun@umich.edu (W.S.)

[†]These authors contributed equally.

[1] Faculty of Engineering, Hokkaido University, Sapporo 060-8628, Japan.
[2] Department of Materials Science and Engineering, UC Berkeley, Berkeley, California 94720, USA
[3] Department of Physics, Tokyo Metropolitan University, Hachioji 192-0397, Japan.
[4] Graduate School of Advanced Science and Engineering, Hiroshima University, 1-3-1 Kagamiyama, Higashihiroshima, 739-8526, Japan
[5] Creative Research Institution Hokkaido University, Kita 21, Nishi 10, Sapporo 001-0021, Japan
[6] Hitachi High-Tech Corporation, Ichige 882 Hitachinaka, 312-8504 Japan
[7] Center for Crystal Science and Technology, University of Yamanashi, Kofu 400-8511, Japan.
[8] Materials Sciences Division, Lawrence Berkeley National Laboratory, Berkeley, CA 94720, USA
[9] Department of Materials Science and Engineering, University of Michigan, Ann Arbor, Michigan, 48109, USA

**Abstract:**

Solid-state synthesis from powder precursors is the primary processing route to advanced multicomponent ceramic materials. Designing ceramic synthesis routes is usually a laborious, trial-and-error process, as heterogeneous mixtures of powder precursors often evolve through a complicated series of reaction intermediates. Here, we show that phase evolution from multiple precursors can be modeled as a sequence of pairwise interfacial reactions, with thermodynamic driving forces that can be efficiently calculated using *ab initio* methods. Using the synthesis of the classic high-temperature superconductor $YBa_2Cu_3O_{6+x}$ (YBCO) as a representative system, we rationalize how replacing the common $BaCO_3$ precursor with $BaO_2$ redirects phase evolution through a kinetically-facile pathway. Our model is validated from *in situ* X-ray diffraction and *in situ* microscopy observations, which show rapid YBCO formation from $BaO_2$ in only 30 minutes. By combining thermodynamic modeling with *in situ* characterization, we introduce a new computable framework to interpret and ultimately design synthesis pathways to complex ceramic materials.

Solid-state ceramic synthesis involves heating a mixture of precursor powders at high temperatures (typically >700 °C) and has been used to realize countless functional materials (*1-3*). Recent *in situ* characterization studies have revealed that solid-state reactions often evolve through a variety of non-equilibrium intermediates prior to formation of the equilibrium phase (*4-10*). These complicated phase evolution sequences are currently difficult to understand, resulting in laborious trial-and-error efforts to optimize ceramic synthesis recipes. The ability to rationalize and anticipate which intermediate phases form would enable solid-state chemists to design crystallization pathways that target (or avoid) specific intermediates, accelerating the design of time- and energy-efficient ceramic synthesis recipes for new materials. The need for synthesis prediction encourages the use of computation as a guide, but computation has mostly been used only to evaluate thermodynamic stability or metastability (*11-15*). While useful, phase stability alone does not provide any mechanistic insights to help guide synthesis planning for experimentalists.

The complexity of phase evolution arises from the various pathways by which an initially heterogeneous mixture of precursor particles can transform to a homogeneous target phase. At the microscopic level, solid-state reactions initiate in the interfacial regions between precursors as the system is heated. Because these interfacial reactions occur between only two solid phases at a time, we hypothesize that reactions starting from three or more precursors can be modeled as being initially be dominated by the most reactive interface between a single pair of precursors, as illustrated schematically in **Figure 1A**. Once two precursors react to form a new phase, this non-equilibrium intermediate will then react through its interface with other precursors and intermediate phases. By decomposing the overall phase evolution into a sequence of pairwise reactions, we can calculate the thermodynamics and analyze the kinetics of each reaction step separately, which provides a simplified theoretical picture to conceptualize and navigate ceramic synthesis (*15-19*).

We demonstrate how this concept of sequential pairwise reactions enables us to model phase evolution in the ceramic synthesis of the classic high-temperature superconductor, $YBa_2Cu_3O_{6+x}$ (YBCO) (*20-22*). Following the discovery that YBCO remains superconducting above the boiling point of liquid $N_2$ (>77 K), YBCO has been synthesized many thousands of times in laboratories around the world. The typical synthesis recipe for YBCO calls for three precursors—a 0.5/2/3 molar ratio of $Y_2O_3$/$BaCO_3$/CuO powders—which are ground in a mortar, then compacted, pelletized, and baked in air at 950 °C for >12 hours. Even after 12 hours, the synthesis reaction is often incomplete, so the pellets must be re-ground, re-pelletized, and re-baked until phase-pure YBCO is obtained (*23*).

It has been reported that replacing $BaCO_3$ with $BaO_2$ can shorten YBCO synthesis times to 4 hours and eliminate the need for regrinding (*24, 25*). This dramatic difference in synthesis times offers an ideal case study to explore how precursor selection governs phase evolution in solid-state synthesis (*26*). In **Figure 1B**, we show temperature-dependent Gibbs reaction energies, $\Delta G_{rxn}$, for the formation of YBCO with either $BaO_2$ or $BaCO_3$ as the barium source. $BaO_2$ is less stable than $BaCO_3$ (*27*), so although both reactions are thermodynamically favorable ($\Delta G_{rxn} < 0$) above ~700 °C, the thermodynamic driving force (magnitude of $\Delta G_{rxn}$) is much larger with $BaO_2$.

Naively, one might anticipate that this larger driving force explains why YBCO synthesis with a $BaO_2$ precursor proceeds faster. Instead, we will show that it is actually the pairwise $BaO_2$|CuO reaction that directs phase evolution through a low-temperature eutectic melt, producing a liquid self-

flux to facilitate rapid YBCO formation. $BaO_2$ is a relatively uncommon YBCO precursor, appearing in only 8 out of 237 synthesis recipes for YBCO (and related phases) as text-mined from the literature (*28*), whereas $BaCO_3$ is the most common Ba precursor, at 176 out of 237 recipes (**Table S1** shows all extracted synthesis recipes). By better understanding how uncommon precursors promote kinetically-facile sequential pairwise reactions (*26*), we can build towards new design principles for precursor selection and rational synthesis planning.

Here, we use *in situ* synchrotron X-ray diffraction (XRD) to characterize the temperature-time-transformation process of YBCO formation, as well as *in situ* microscopy (SEM, DF-STEM) to directly observe the spatiotemporal microstructural evolution from the three initial precursors. By comparing these experimentally-observed phase evolution pathways against density functional theory (DFT)-calculated thermodynamics (*29*) aided by a machine-learned model for temperature-dependent Gibbs free energies (*30*), we conclusively verify the role of interfacial reactions in dictating phase evolution in solid-state synthesis. Our model provides a theoretical foundation to model phase evolution during solid-state synthesis from multiple precursors, integrating computation and experiment towards the long-standing goal of predictive solid-state synthesis.

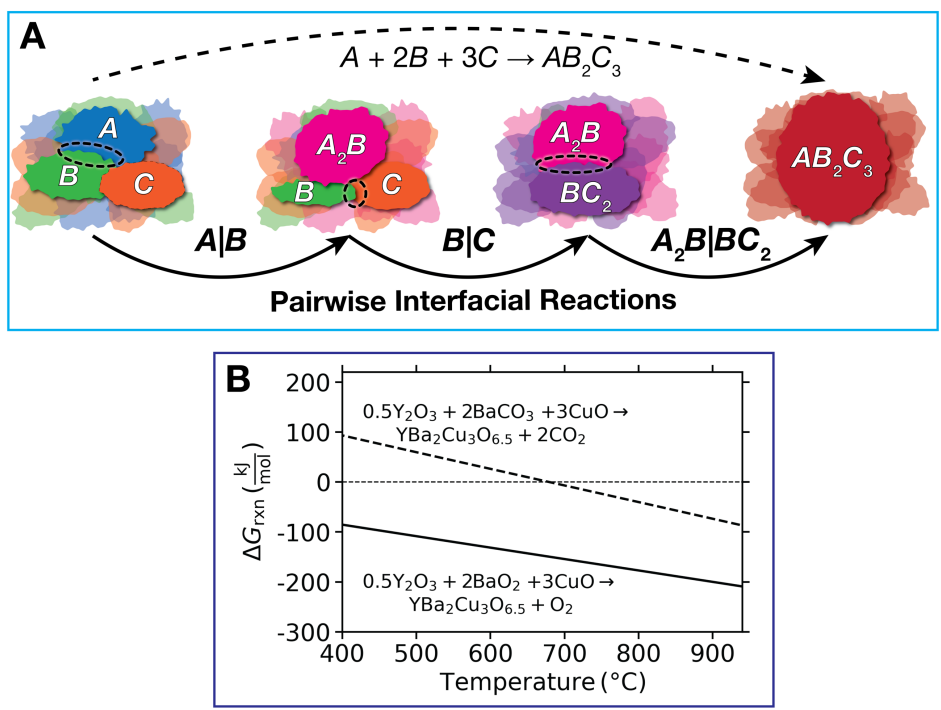

**Figure 1. Schematic of sequential pairwise interfacial reactions and overall reaction energetics for YBCO synthesis.** (**A**) Schematic of the pairwise reaction concept, illustrating that phase evolution from powder precursors must initiate at the shared interface between two precursor grains. (**B**) The temperature-dependent Gibbs reaction energies, $\Delta G_{rxn}$, for the formation of YBCO from precursor mixtures utilizing $BaCO_3$ (dashed line) or $BaO_2$ (solid line) as the Ba source.

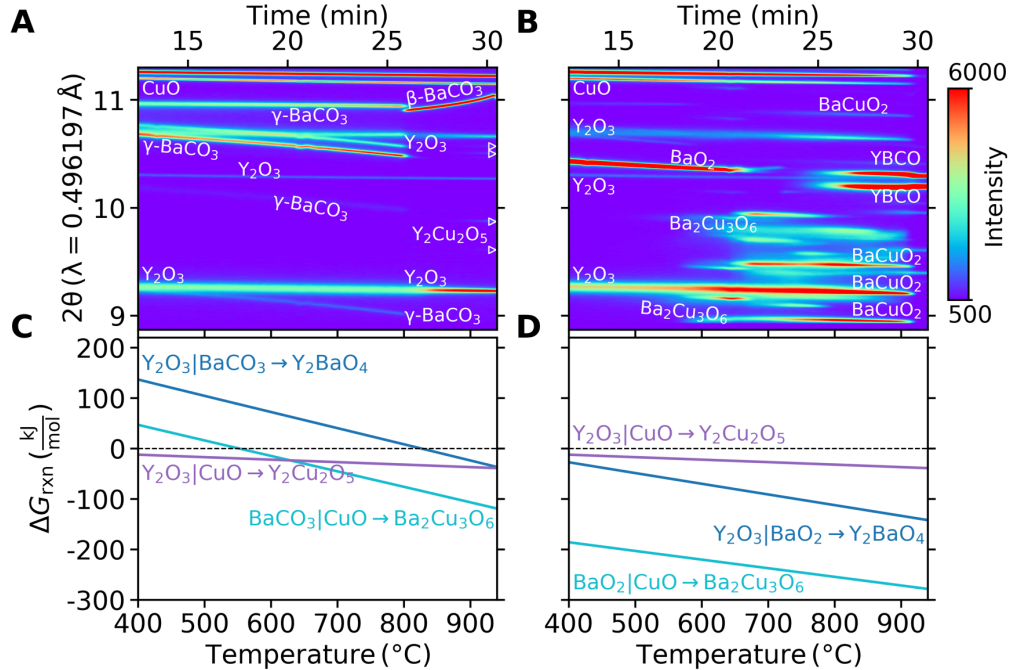

**Figure 2. Phase evolution during YBCO synthesis compared to reaction thermodynamics. (A)** *in situ* synchrotron XRD pattern for heating of the $Y_2O_3$ + $BaCO_3$ + $CuO$ precursor mixture. The triangles mark peaks for $Y_2Cu_2O_5$. Individual XRD patterns at select temperatures are provided in **Supplementary Figure 1**. **(B)** *in situ* synchrotron XRD pattern for heating of the $Y_2O_3$ + $BaO_2$ + $CuO$ precursor mixture. Individual XRD patterns at select temperatures are provided in **Supplementary Figure 2**. Mixed powders were heated in quartz tubes under air atmosphere at a heating rate of 30 °C/min. **(C)** Gibbs reaction energies for the lowest energy reactions at each interface in the $Y_2O_3$ + $BaCO_3$ + $CuO$ precursor mixture. The reactions are $Y_2O_3|BaCO_3$ = 1.5 $Y_2O_3$ + 1.5 $BaCO_3$ → 1.5 $BaY_2O_4$ + 1.5 $CO_2$; $Y_2O_3|CuO$ = 1.5 $Y_2O_3$ + 3 $CuO$ → 1.5 $Y_2Cu_2O_5$; $BaCO_3|CuO$ = 12/7 $BaCO_3$ + 18/7 $CuO$ + 3/7 $O_2$ → 6/7 $Ba_2Cu_3O_6$ + 12/7 $CO_2$, **(D)** Gibbs reaction energies for the lowest energy reactions at each interface in the $Y_2O_3$ + $BaO_2$ + $CuO$ precursor mixture. The reactions are $Y_2O_3|BaO_2$ = 2 $Y_2O_3$ + 2 $BaO_2$ → 2 $BaY_2O_4$ + $O_2$, $Y_2O_3|CuO$ = 1.5 $Y_2O_3$ + 3 $CuO$ → 1.5 $Y_2Cu_2O_5$, $BaO_2|CuO$ = 2.4 $BaO_2$ + 3.6 $CuO$ → 1.2 $Ba_2Cu_3O_6$ + 0.6 $O_2$. The coefficients of each reaction are normalized to be consistent with the formation of 1 mol of $YBa_2Cu_3O_{6.5}$ in an atmosphere open to $O_2$. As such, the products of each reaction form 6 mol of non-oxygen atoms. See the **Supplementary Information** for more details.

In **Figure 2**, we show *in situ* synchrotron X-ray diffraction patterns for phase evolution in YBCO synthesis in air with either $BaCO_3$ (**Figure 2A**) or $BaO_2$ (**Figure 2B**) as the Ba source, which we compare to the thermodynamic driving force for new phase formation at each pairwise interface (**Figure 2C-D**). **Figure 2A** shows that when $BaCO_3$ is used, the precursors remain the dominant phases up to 940 °C, confirming the lack of rapid phase formation. In contrast, **Figure 2B** shows the formation of YBCO in 30 min when $BaO_2$ is used as the Ba source. In both cases, we have a three-precursor system, so the relevant interfaces are $Y_2O_3|CuO$, $Y_2O_3|BaCO_3(BaO_2)$, and $BaCO_3(BaO_2)|CuO$. In the $BaCO_3$-containing system, no reaction has substantial driving force until >900 °C (**Figure 2C**). When $BaCO_3$ is replaced with $BaO_2$, the reaction thermodynamics change dramatically as the $BaO_2|CuO$ interface has large driving force ($\Delta G_{rxn} < -200$ kJ/mol) to form ternary Ba-Cu-oxides above 400 °C (**Figure 2D**). This is consistent with *in situ* XRD observations of barium copper oxides emerging at ~600 °C and the consumption of $BaO_2$ by ~700 °C (**Figure 2B**).

Synthesis of YBCO using a $BaCO_3$ precursor usually requires >12 hours with intermittent re-grindings (*23*), so it is not surprising that YBCO did not form in our 30 min *in situ* experiment (**Figure 2A**). At temperatures >850 °C, traces of a $Y_2Cu_2O_5$ phase are observed, even though the $BaCO_3|CuO$ interface has the larger thermodynamic driving force to react (**Figure 2C**). $BaCO_3$ decomposition is reported to have a substantial activation barrier of 305 kJ/mol (*31*), and the thermodynamic driving forces for all $Y_2O_3$-$BaCO_3$-$CuO$ interfacial reactions have $\Delta G_{rxn}$ less negative than −100 kJ/mol up to 800 °C, which is evidently too small to overcome this kinetic barrier. These poor reaction kinetics, coupled with a small thermodynamic driving force, underlie the slow synthesis of YBCO when starting from a $BaCO_3$ precursor.

The fast formation of YBCO when starting from $BaO_2$ originates from the large driving force at the $BaO_2|CuO$ interface, which is ~200 kJ/mol larger than at the $BaCO_3|CuO$ interface at 600 °C. We previously demonstrated in the $Na_xMO_2$ (M = Co, Mn) system that the first phase to form in an interfacial reaction is the compound with the largest compositionally-unconstrained reaction energy from the precursors (*10*). Here, we confirm this theory in the YBCO system. We calculate that $Ba_2Cu_3O_6$ has the largest reaction energy to form at the $BaO_2|CuO$ interface, and indeed this is the first observed ternary phase, which is accompanied by evolution of $O_2$ gas. Between 600 °C and 850 °C, $Ba_2Cu_3O_6$ decomposes to form $BaCuO_2$ and CuO (**Figure 2B**). The preferential reactivity of the $BaO_2|CuO$ interface—instead of the $Y_2O_3|BaO_2$ or $Y_2O_3|CuO$ interfaces—supports the theory that the first phase to form is the one with the largest thermodynamic driving force, and further asserts that when multiple competing interfaces exist, the interface with the most exergonic compositionally-unconstrained reaction energy will initiate the solid-state reaction. This provides a straightforward means by which computation can be integrated into synthesis planning to predict which pairwise interfaces will be the most reactive in a given precursor mixture and which phases will be most likely to form at those interfaces.

Whereas *in situ* XRD measurements track the temperature-time-transformation evolution of the system, *in situ* SEM/DF-STEM provides direct spatiotemporal observation of the microstructural evolution during the solid-state reaction. We monitored the synthesis of YBCO from $Y_2O_3$ + $BaO_2$ + CuO on a hot stage using *in situ* electron microscopy (SEM/DF-STEM: Hitachi HF5000). Although the *in situ* microscopy used here cannot identify crystal structure, the reaction conditions (temperature, heating rate, precursors) are the same as those characterized by *in situ* XRD (**Figure 2B**), so we anticipate that the temperature-time-transformation progression between the two methods are comparable. We also characterize the elemental distribution in the sample using energy-dispersive X-ray spectroscopy (EDX) before and after the *in situ* microscopy experiment (our EDX instrument can only operate at room temperature). In **Figure 3A**, we show DF-STEM snapshots of the particles during heating along with EDX before and after heating. A video of this reaction is also provided as **Supplementary File 1**.

At room temperature, EDX shows that the three precursor powders are in intimate contact. Importantly, it is clear from EDX that all three potential pairwise interfaces ($Y_2O_3|BaO_2$, $Y_2O_3|CuO$, and $BaO_2|CuO$) exist in the sample. As the stage is heated to 500 °C, the initial $BaO_2$ and CuO precursors react at the $BaO_2|CuO$ interface, which according to the *in situ* XRD experiments, results in $Ba_2Cu_3O_6$. Meanwhile, the $Y_2O_3$ particle remains inert, as does its interface with $BaO_2$. From 650 °C to 800 °C, we observe the ejection of small bubble-like particles, which corresponds to the reaction: $Ba_2Cu_3O_6 \rightarrow 2\ BaCuO_2 + CuO + 0.5\ O_2$. In a separate *in situ* heating experiment, we confirm with SEM and EDX measurements that this initial reaction occurs strictly in the Ba-Cu-O subsystem. (**Supplementary Figure 3**). The observed reactivity of the $BaO_2|CuO$ interface and inertness of the $Y_2O_3$-containing interface aligns with our thermodynamic predictions from **Figure 2D**.

From **Figure 1B**, we calculated the total thermodynamic driving force of $0.5\ Y_2O_3 + 2\ BaO_2 + 3\ CuO \rightarrow YBa_2Cu_3O_{6.5} + O_2$ to be approximately −200 kJ/mol. For the formation of $BaCuO_2$, we calculate a reaction energy of −130 kJ/mol ($2\ BaO_2 + 2\ CuO \rightarrow 2\ BaCuO_2 + O_2$), meaning that ~2/3$^{rd}$ of total reaction driving force is consumed before $Y_2O_3$ becomes involved in the reaction. Only ~70 kJ/mol remain to drive the reaction to form YBCO. This is more or less comparable to the overall reaction energy from $Y_2O_3$, $BaCO_3$ and CuO (**Figure 1B**), indicating this thermodynamic driving force does not account for the quick formation of YBCO when $BaO_2$ is used. Thus, we anticipate kinetic selection to play the primary role in the formation of the next phase. Indeed, this kinetic mechanism is provided by the melting of $BaCuO_2$ and CuO at ~900 °C. This liquid Ba-Cu-O melt is then rapidly consumed into the $Y_2O_3$ particle to form YBCO. In the EDX taken after the experiment, the morphology of the Y region remains similar to the beginning of the experiment, but now Ba and Cu signals are found in the final particle.

In **Figure 3B**, we overlay the observed phase evolution sequence onto the pseudo-binary $BaO_2$-CuO slice (*32*) of the overall $Y_2O_3$-$BaO_2$-CuO phase diagram to reveal how the $BaO_2$ precursor enables rapid YBCO synthesis. The first reaction occurs before 500 °C and proceeds at the $BaO_2|CuO$ interface to form $Ba_2Cu_3O_6$. This is consistent with our calculations in **Figure 2D**, where we found the $BaO_2|CuO$ interface to be the most reactive among the three precursor interfaces and $Ba_2Cu_3O_6$ to be the phase with the largest driving force to form at this interface. Above 700°C, $Ba_2Cu_3O_6$ undergoes peritectoid decomposition into $BaCuO_2$ and CuO, which was observed as the ejection of small bubble-like particles in **Figure 3**. $BaCuO_2$ and CuO are unreactive until the temperature is increased to their eutectic point at 890 °C, after which $BaCuO_2$ and CuO melt into one another. This liquid melt becomes a self-flux, providing fast kinetic transport of Ba and Cu into $Y_2O_3$ for the rapid formation of YBCO at the $Y_2O_3|Ba$-Cu-O(liquid) interface.

If one consults the $Y_2O_3$-CuO or $Y_2O_3$-$BaO_2$ phase diagrams (*33*), the lowest liquidus temperatures in these systems are ≥ 1095 °C, which is above the temperature at which YBCO decomposes (1006 °C) (*34*). $BaO_2$ therefore plays a crucial role in directing the phase evolution through the pseudo-binary $BaO_2$-CuO subsystem—where a low-temperature liquid self-flux provides the fast diffusion kinetics that facilitates the formation of YBCO in 30 minutes. This is in contrast to when $BaCO_3$ is used as the Ba source, where the slow decomposition reaction kinetics at the $BaCO_3|CuO$ interface forces the overall reaction to proceed through the $Y_2O_3$-CuO subsystem, and a high liquidus temperature of 1095 °C obstructs any liquid-mediated transport kinetics for YBCO formation (*33*).

Although the overall reaction energies shown in **Figure 1B** suggest that the larger thermodynamic driving force is why a reaction with the $BaO_2$ precursor proceeds more quickly than with $BaCO_3$, we emphasize here that the magnitude of the overall reaction energy is not the origin of the fast synthesis time. Instead, it is the initial selection of the $BaO_2$-CuO subsystem, where there is a low-temperature eutectic below the decomposition temperature of YBCO, that enables rapid YBCO synthesis by forming a self-flux. This finding highlights the need to consider computations beyond the phase stabilities of the target or overall reaction energies in order to obtain mechanistic insights into the reaction pathways by which phases can evolve during synthesis.

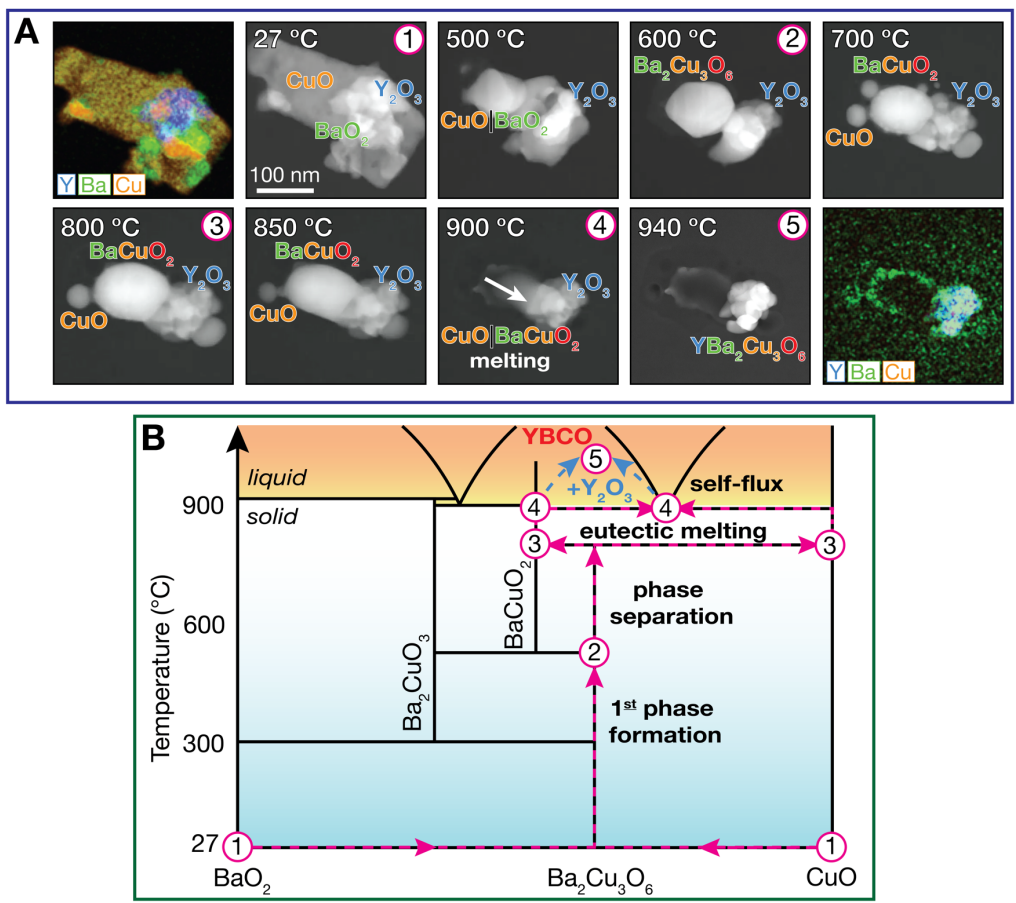

**Figure 3.** *In situ* microscopy of YBCO formation from $Y_2O_3$, $BaO_2$, and CuO particles and the observed phase evolution sequence mapped onto the $BaO_2$-CuO phase diagram. **(A)** *In situ* DF-STEM and EDX images show the heating of 0.5 $Y_2O_3$ + 2 $BaO_2$ + 3 CuO from 27 °C to 940 °C at 30 °C/min. The markers in the upper right corner of select panels are for comparison to panel B. A video of the reaction is provided in Supplementary File 1. *In situ* SEM and EDX for a shorter run to capture the initial formation of $Ba_2Cu_3O_6$ is also provided in Supplementary Figure 3. **(B)** Observed phase evolution sequence in the context of the pseudo-binary phase diagram for $BaO_2$-CuO, adapted from Ref. (*32*).

Upon cooling the sample down from 940 °C at a rate of 5 °C/min, *in situ* XRD shows in **Figure 4** a structural transition from tetragonal to orthorhombic YBCO at 620 °C, indicating the uptake of ambient $O_2$ into $YBa_2Cu_3O_6$ to form $YBa_2Cu_3O_{6+x}$, consistent with reports from the literature (*35, 36*).

The synthesized product exhibits a strong diamagnetic signal below 77 K (**Figure 4C**), indicating the successful synthesis of superconducting YBCO. From a thermodynamic perspective, it is well-characterized that $YBa_2Cu_3O_{6+x}$ is metastable at low temperature with respect to decomposition (*37*) by the reaction:

$$YBa_2Cu_3O_{6.5} + 0.5\ O_2 \rightarrow 0.5\ Y_2O_3 + 1\ Ba_2Cu_3O_6 \qquad \Delta G_{rxn} \approx -100\ kJ/mol\ at\ 27\ °C$$

However, this solid-state decomposition is kinetically limited at room temperature. On the other hand, oxygen diffusion is highly mobile in the YBCO framework (*38, 39*), indicating that this final topotactic uptake of $O_2$ gas at the YBCO|$O_2$ interface is a kinetically-mediated non-equilibrium reaction.

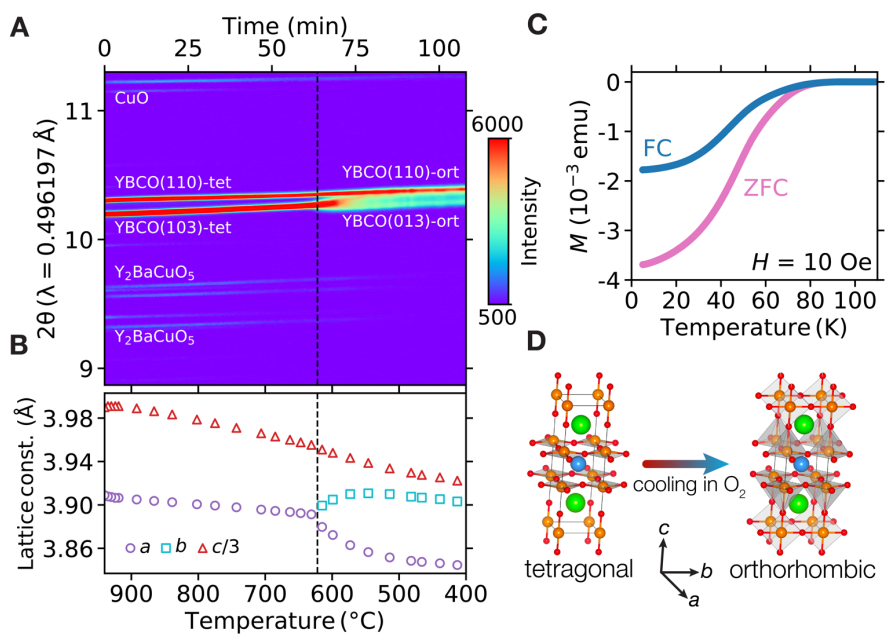

**Figure 4. Topotactic $O_2$ uptake and phase transition during slow cooling.** (**A**) *in situ* synchrotron XRD pattern for cooling of $Y_2O_3$ + $BaO_2$ + CuO precursor from 940 °C to 400 °C at 5 °C/min. "tet" refers to the tetragonal structure and "ort" to the orthorhombic structure. (**B**) Changes in lattice parameters during cooling. (**C**) Magnetic susceptibility of synthesized YBCO exhibiting superconductivity above liquidus nitrogen temperature. (**D**) The tetragonal and orthorhombic crystal structures for YBCO, where blue spheres are Y, green are Ba, orange are Cu, and red are O.

In **Figure 5**, we summarize how phase evolution during YBCO synthesis can be understood as a sequence of pairwise reactions that result from an interplay between thermodynamics and kinetics. The initial mixture of three precursors—$Y_2O_3$, $BaO_2$ and CuO—produces three possible reactive interfaces. We calculated in **Figure 2D** that the $BaO_2$|CuO interface possesses the largest thermodynamic driving force to react, and predicted $Ba_2Cu_3O_6$ to be the first reaction intermediate, which was confirmed by *in situ* XRD (**Figure 2B**) and microscopy (**Figure 3A**, **Supplementary Figure 3**). The formation of $Ba_2Cu_3O_6$ below 600 °C consumes ~2/3$^{rd}$ of the overall reaction driving force, meaning the ensuing reactions necessarily occur with smaller driving forces. Using *in situ* DF-STEM we observed that after the peritectoid decomposition of $Ba_2Cu_3O_6$ into $BaCuO_2$ + CuO, there is no further phase evolution in the system until the formation of a eutectic melt at the $BaCuO_2$|CuO interface. This liquid melt serves as a self-flux, providing fast Ba and Cu transport into the thus-far immobile $Y_2O_3$, forming $YBa_2Cu_3O_6$ (**Figure 3**). Finally, fast topotactic oxygen uptake at the $YBa_2Cu_3O_6$|$O_2$ interface upon cooling yields the superconducting $YBa_2Cu_3O_{6+x}$ phase (**Figure 4**), which persists kinetically as a metastable phase to room temperature, instead of decomposing to the equilibrium $Y_2O_3$ + $Ba_2Cu_3O_6$ phases. In the $Y_2O_3$-$BaCO_3$-CuO precursor set, small $BaCO_3$|CuO reaction driving forces and poor $BaCO_3$ decomposition kinetics drive the phase evolution down through the $Y_2O_3$-CuO subsystem, where slow diffusion kinetics means manual regrinding is necessary to reintroduce interfaces between any unfinished reaction intermediates.

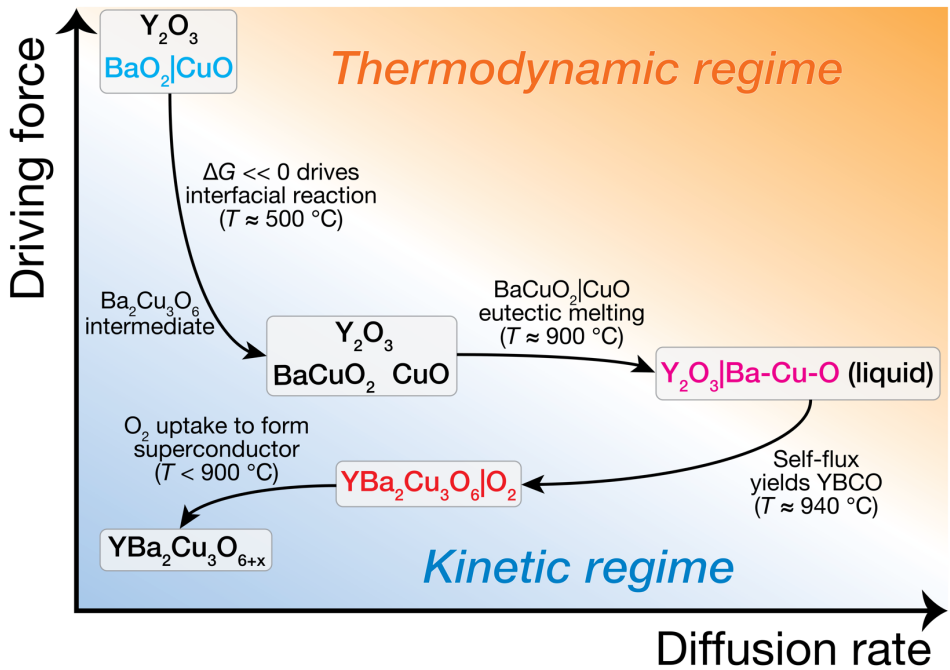

**Figure 5.** Phase evolution pathway for the formation of YBCO dictated by sequential pairwise reactions. The YBCO synthesis pathway is shown here along two qualitative axes – the thermodynamic driving force to form new phases along the vertical axis and the diffusion rate of reactive species along the horizontal axis. Within this framework, we understand reaction events occurring in either a thermodynamic regime, where driving forces or diffusion rates are sufficiently high that equilibrium products are observed, or a kinetic regime, where ion transport is sufficiently slow or driving forces sufficiently small such that the system becomes unreactive or non-equilibrium products are formed.

Our investigation here provides a general conceptual framework to understand the solid-state synthesis of complex multicomponent ceramics. A ceramic synthesis reaction that begins from $N$ precursors will exhibit $_NC_2$ pairwise reaction interfaces. We showed here that of the many possible pairwise interfaces, the first reaction will occur between the two precursors with the largest compositionally-unconstrained reaction driving force. This initial reaction interface can be predicted from *ab initio* thermodynamics and determines which pseudo-binary subsystem the ensuing phase evolution proceeds from. By thoughtfully choosing the starting precursors (*26*) to control which pairwise interface is the most reactive, one can deliberately direct phase evolution through whichever pseudo-binary subsystem exhibits the best kinetic pathway to the target material. Today, it remains difficult to anticipate which kinetic mechanisms are available in a given subsystem, meaning that in the short term, *in situ* characterization will be the most productive approach for rationally designing solid-state synthesis recipes. In the future, a theoretical framework that embeds nucleation, diffusion, and crystal growth kinetics within a thermodynamic description of sequential pairwise reactions will pave the way towards a complete computational platform for predictive solid-state ceramic synthesis.

**Materials and Methods**
**In-situ synchrotron powder X-ray diffraction**

$Y_2O_3$ (>99.9%, Kojundo Kagaku), $BaCO_3$ (>99.9 %, Kojyundo Kagaku), $BaO_2$ (>80%, Jyunsei Kagaku), CuO (>99%, Wako Chemical) were weighed in a molar ratio of for Y/Ba/Cu =1/2/3, and loaded into a zirconia pot with zirconia balls with a diameter of 4 mm. The starting materials were mechanochemically milled by planetary ball milling for 3 h over 150 rpm. The mixed powder was loaded into a quartz capillary with a diameter of 0.3 mm.

The change in crystalline phases were examined using synchrotron powder X-ray diffraction at the *BL02B2* beamline of *SPring-8* (proposal numbers 2019A1101, 2019B1195 and 2020A1096). The quartz capillary with powder mixture was settled in a furnace in air atmosphere. Heating started after setting in the furnace operated at 100 °C at the heating rate of 30 °C /min till 940 °C. The sample kept 10 min at 940 °C and then started cooling at 5 °C /min till 400 °C. The diffraction data of $2\theta$ range from 8.9° to 15.5° with a step of 0.02° were collected using a high-resolution one-dimensional semiconductor detector (MYTHEN)(*40*). The wavelength of the radiation beam was determined using a $CeO_2$ standard. The crystal structure was visualized using VESTA software.(*41*)

**In-situ TEM measurement**

In an Ar-filled glove box, $BaO_2$ powder (>80%, Jyunsei Kagaku) was mechanochemically milled by planetary ball milling for 8 h over 150 rpm. The powder was sieved to remove particles larger than 20 μm. In ambient atmosphere, $Y_2O_3$ (>99.9%, Kojundo Kagaku), CuO nanopowder (>99%, Alderich), and above $BaO_2$ powder were weighed in a molar ratio of for Y/Ba/Cu =1/2/3, and loaded again into a zirconia pot with φ 4 mm zirconia balls. The powder was mechanochemically mixed by planetary ball milling for 3 h over 150 rpm. The sample was dispersed in dehydrate ethanol, and ultrasonicated. This suspension was dropped onto a silicon nitride TEM grid.

Morphology and compositional change were observed by transmission electron microscopy (TEM: HF-5000 Hitachi High-Tech Corporation). The accelerate voltage was 200 kV, and pressure was approximately $2\times10^{-5}$ Pa. The sample was initially heated at 300 °C, and then heated till 940 °C at 30 °C /min. The apparatus allows to record three images simultaneously: scanning electron microscope (SEM), bright-field scanning transmission electron microscopy (BF-STEM), and dark-field scanning transmission electron microscopy (DF-STEM) images. Before and after heating the sample, compositional distribution was examined by energy-dispersive X-ray (EDX) mapping at room temperature.

**Magnetization measurement**

Magnetization was measured using a superconducting quantum interference device (SQUID) magnetometer (Quantum Design MPMS-3) with an applied field of 10 *Oe*, in order to check Meissner effect of synthesized sample.

**Computational**

Standard Gibbs formation energies, $\Delta G°_f(T)$, for gaseous species were obtained from NIST (*42*). To account for the synthesis atmosphere (air), Gibbs formation energies of a given gaseous species, $\Delta G°_{f,i}(T)$, were obtained as:

$$\Delta G_{f,i}(T) = \Delta G^o_{f,i}(T) + RTln(p_i)$$

where R is the gas constant and $p_i$ approximates the activity coefficient of gaseous species, *i*. The only gaseous species evolved or consumed in reactions discussed in this work are $O_2$ and $CO_2$, where $p_{O2}$ was taken to be 0.21 atm and $p_{CO2}$ = 0.0004 atm.

For solid-state compounds, formation enthalpies (at 0 K) were obtained with density functional theory (DFT), utilizing the SCAN meta-GGA density functional (*29*). Each structure was obtained from the Materials Project database (*43*) and optimized using the Vienna Ab Initio Simulation Package (VASP) (*44*) and the projector augmented wave (PAW) method (*45*), a plane-wave energy cutoff of 520 eV, and 1000 k-points per reciprocal atom.

Standard Gibbs formation energies, $\Delta G°_f(T)$, for each solid-state compound were then obtained by combining the DFT-calculated formation enthalpies, the machine-learned descriptor introduced in (*30*), and experimental Gibbs energy data for elemental phases as described in (*30*). The activity of all solid phases was taken to be 1, so $\Delta G_f(T) = \Delta G°_f(T)$.

Gibbs reaction energies, $\Delta G_{rxn}(T)$ were obtained as:

$$\Delta G_{rxn}(T) = \sum_{products} \Delta G_f(T) - \sum_{reactants} \Delta G_f(T)$$

The coefficients of each reaction were selected such that 6 moles of non-oxygen atoms appear in the product side of each reaction. This was done to normalize the comparison of $\Delta G_{rxn}(T)$ across a diverse set of reactions, and because the reacting mixture was assumed to exchange freely with $O_2$ in the synthesis atmosphere.

**References**
1. W. D. Kingery, H. K. Bowen, D. R. Uhlmann, *Introduction to Ceramics, 2nd Edition*. (Wiley-Blackwell, Hoboken, New Jersey, 1976).
2. M. G. Kanatzidis *et al.*, Report from the third workshop on future directions of solid-state chemistry: The status of solid-state chemistry and its impact in the physical sciences. *Prog. Solid State Chem.* **36**, 1-133 (2008).
3. A. R. West, *Solid State Chemistry and its Applications*. (Wiley, 2014).
4. M. H. Nielsen, S. Aloni, J. J. De Yoreo, In situ TEM imaging of $CaCO_3$ nucleation reveals coexistence of direct and indirect pathways. *Science* **345**, 1158-1162 (2014).
5. A. J. Martinolich, J. A. Kurzman, J. R. Neilson, Circumventing Diffusion in Kinetically Controlled Solid-State Metathesis Reactions. *J. Am. Chem. Soc.* **138**, 11031-11037 (2016).
6. Z. Jiang, A. Ramanathan, D. P. Shoemaker, In situ identification of kinetic factors that expedite inorganic crystal formation and discovery. *J. Mater. Chem. C* **5**, 5709-5717 (2017).
7. A. S. Haynes, C. C. Stoumpos, H. Chen, D. Chica, M. G. Kanatzidis, Panoramic Synthesis as an Effective Materials Discovery Tool: The System Cs/Sn/P/Se as a Test Case. *J. Am. Chem. Soc.* **139**, 10814-10821 (2017).
8. H. He *et al.*, Combined computational and experimental investigation of the $La_2CuO_{4-x}S_x$ ($0 \leq x \leq 4$) quaternary system. *PNAS* **115**, 7890-7895 (2018).
9. H. Kohlmann, Looking into the Black Box of Solid-State Synthesis. *Eur. J. Inorg. Chem.* **2019**, 4174-4180 (2019).
10. M. Bianchini *et al.*, The interplay between thermodynamics and kinetics in the solid-state synthesis of layered oxides. *Nat Mater*, (2020).
11. S. P. Ong, L. Wang, B. Kang, G. Ceder, Li−Fe−P−O2 Phase Diagram from First Principles Calculations. *Chem. Mater.* **20**, 1798-1807 (2008).
12. W. Sun *et al.*, The thermodynamic scale of inorganic crystalline metastability. *Sci. Adv.* **2**, e1600225-e1600225 (2016).
13. C. J. Bartel, A. W. Weimer, S. Lany, C. B. Musgrave, A. M. Holder, The role of decomposition reactions in assessing first-principles predictions of solid stability. *npj Computational Materials* **5**, 4 (2019).
14. A. Narayan *et al.*, Computational and experimental investigation for new transition metal selenides and sulfides: The importance of experimental verification for stability. *Phys. Rev. B* **94**, (2016).
15. M. Jansen, A Concept for Synthesis Planning in Solid-State Chemistry. *Angew. Chem. Int. Ed.* **41**, 3746-3766 (2002).
16. F. J. DiSalvo, Solid-State Chemistry: A A Rediscovered Chemical Frontier. *Science* **247**, 649-655 (1990).
17. A. Sleight, *Synthesis of Oxide Superconductors*. (1991), vol. 44, pp. 24-30.
18. J. R. Chamorro, T. M. McQueen, Progress toward Solid State Synthesis by Design. *Acc. Chem. Res.* **51**, 2918-2925 (2018).
19. D. L. M. Cordova, D. C. Johnson, Synthesis of Metastable Inorganic Solids with Extended Structures. *ChemPhysChem* **21**, 1345-1368 (2020).
20. M. Wu *et al.*, Superconductivity at 93 K in a new mixed-phase Y-Ba-Cu-O compound system at ambient pressure. *Phys. Rev. Lett.* **58**, 908-910 (1987).
21. S. Hikami, T. Hirai, S. Kagoshima, High Transition Temperature Superconductor: Y-Ba-Cu Oxide. *Jpn. J. Appl. Phys.* **26**, L314-L315 (1987).

22. R. J. Cava et al., Bulk superconductivity at 91 K in single-phase oxygen-deficient perovskite Ba$_2$YCu$_3$O$_{9-\delta}$. *Phys. Rev. Lett.* **58**, 1676-1679 (1987).
23. P. Grant, in *New Scientist*. (New Scientist Ltd., London, 1987), chap. 36, pp. 36-39.
24. C. A. Costa et al., Synthesis of YBa$_2$Cu$_3$O$_{7-x}$ polycrystalline superconductors from Ba peroxide: First physico-chemical characterization. *J. Cryst. Growth* **85**, 623-627 (1987).
25. B. D. Fahlman, Superconductor Synthesis—An Improvement. *J. Chem. Educ.* **78**, 1182 (2001).
26. X. Jia et al., Anthropogenic biases in chemical reaction data hinder exploratory inorganic synthesis. *Nature* **573**, 251-255 (2019).
27. J. L. Jorda, T. K. Jondo, Barium oxides: equilibrium and decomposition of BaO$_2$. *J. Alloys Compd.* **327**, 167-177 (2001).
28. O. Kononova et al., Text-mined dataset of inorganic materials synthesis recipes. *Scientific Data* **6**, 203 (2019).
29. J. Sun, A. Ruzsinszky, J. P. Perdew, Strongly Constrained and Appropriately Normed Semilocal Density Functional. *Phys. Rev. Lett.* **115**, 036402 (2015).
30. C. J. Bartel et al., Physical descriptor for the Gibbs energy of inorganic crystalline solids and temperature-dependent materials chemistry. *Nat Commun* **9**, 4168 (2018).
31. I. Arvanitidis, D. Siche, S. Seetharaman, A study of the thermal decomposition of BaCO$_3$. *Metall. Mater. Trans. B* **27**, 409-416 (1996).
32. G. F. Voronin, S. A. Degterov, Solid State Equilibria in the Ba-Cu-O System. *J. Solid State Chem.* **110**, 50-57 (1994).
33. B.-J. Lee, D. N. Lee, Thermodynamic Evaluation for the Y$_2$O$_3$–BaO–CuO$_x$ System. *J. Am. Ceram. Soc.* **74**, 78-84 (1991).
34. W. Wong-Ng, L. P. Cook, Liquidus Diagram of the Ba-Y-Cu-O System in the Vicinity of the Ba$_2$YCu$_3$O$_{6+x}$ Phase Field. *J. Res. Natl. Inst. Stand. Technol.* **103**, 379-403 (1998).
35. J. D. Jorgensen et al., Oxygen ordering and the orthorhombic-to-tetragonal phase transition in YBa$_2$Cu$_3$O$_{7-x}$. *Phys. Rev. B* **36**, 3608-3616 (1987).
36. R. P. Stoffel, C. Wessel, M. W. Lumey, R. Dronskowski, Ab initio thermochemistry of solid-state materials. *Angew. Chem. Int. Ed. Engl.* **49**, 5242-5266 (2010).
37. E. L. Brosha, P. K. Davies, F. H. Garzon, I. D. Raistrick, Metastability of Superconducting Compounds in the Y-Ba-Cu-O System. *Science* **260**, 196-198 (1993).
38. N. Chen, S. J. Rothman, J. L. Routbort, K. C. Goretta, Tracer diffusion of Ba and Y in YBa$_2$Cu$_3$O$_x$. *J. Mater. Res.* **7**, 2308-2316 (1992).
39. D. de Fontaine, G. Ceder, M. Asta, Low-temperature long-range oxygen order in YBa$_2$Cu$_3$O$_z$. *Nature* **343**, 544-546 (1990).
40. S. Kawaguchi et al., High-throughput powder diffraction measurement system consisting of multiple MYTHEN detectors at beamline BL02B2 of SPring-8. *Rev. Sci. Instrum.* **88**, 085111 (2017).
41. K. Momma, F. Izumi, VESTA: a three-dimensional visualization system for electronic and structural analysis. *J. Appl. Crystallogr.* **41**, 653-658 (2008).
42. M. W. Chase, *NIST-JANAF Thermochemical Tables, 4th Edition*. (American Institute of Physics, 1998).
43. A. Jain et al., Commentary: The Materials Project: A materials genome approach to accelerating materials innovation. *APL Materials* **1**, 011002 (2013).
44. G. Kresse, J. Hafner, Ab initio molecular dynamics for liquid metals. *Phys. Rev. B* **47**, 558-561 (1993).

45. G. Kresse, D. Joubert, From ultrasoft pseudopotentials to the projector augmented-wave method. *Phys. Rev. B* **59**, 1758-1775 (1999).

**Acknowledgments:** AM and GY thanks Dr. S. Kawaguchi (JASRI) for technical support for in-situ synchrotron measurement in *SPring-8* with the approvals of 2019A1101, 2019B1195 and 2020A1096. Preliminary TEM observation was performed at the "Joint-use Facilities: Laboratory of Nano-Micro Material Analysis" in Hokkaido University. This work also used computational resources sponsored by the Department of Energy's Office of Energy Efficiency and Renewable Energy, located at NREL. **Funding:** The work by WS was supported by the U.S. Department of Energy (DOE), Office of Science, Basic Energy Sciences (BES), under Award #DE-SC0021130. The computational thermodynamics was supported as part of GENESIS: A Next Generation Synthesis Center, an Energy Frontier Research Center funded by the U.S. Department of Energy, Office of Science, Basic Energy Sciences under Award Number DESC0019212. The experimental work was partially supported by KAKENHI Grant Numbers JP16K21724, JP19H04682 and JP20KK0124.

**Supplementary Materials:**

Supplementary Figures S1-3

Supplementary Table S1

Supplementary File S1

# Supplementary information

# Observing and modeling the sequential pairwise reactions that drive solid-state ceramic synthesis

Akira Miura,[1,*,‖] Christopher J. Bartel,[2,‖] Yusuke Goto,[3] Yoshikazu Mizuguchi,[3] Chikako Moriyoshi,[4] Yoshihiro Kuroiwa,[4] Yongming Wang,[5] Toshie Yaguchi,[6] Manabu Shirai,[6] Masanori Nagao,[7] Nataly Carolina Rosero-Navarro,[1] Kiyoharu Tadanaga,[1] Gerbrand Ceder,[2,8] Wenhao Sun[9,*]

*Correspondence to: amiura@eng.hokudai.ac.jp (A.M.), whsun@umich.edu (W.S.)

‖These authors contributed equally.

[1] Faculty of Engineering, Hokkaido University, Sapporo 060-8628, Japan.

[2] Department of Materials Science and Engineering, UC Berkeley, Berkeley, California 94720, USA

[3] Department of Physics, Tokyo Metropolitan University, Hachioji 192-0397, Japan.

[4] Graduate School of Advanced Science and Engineering, Hiroshima University, 1-3-1 Kagamiyama, Higashihiroshima, 739-8526, Japan

[5] Creative Research Institution Hokkaido University, Kita 21, Nishi 10, Sapporo 001-0021, Japan

[6] Hitachi High-Tech Corporation, Ichige 882 Hitachinaka, 312-8504 Japan

[7] Center for Crystal Science and Technology, University of Yamanashi, Kofu 400-8511, Japan.

[8] Materials Sciences Division, Lawrence Berkeley National Laboratory, Berkeley, CA 94720, USA

[9] Department of Materials Science and Engineering, University of Michigan, Ann Arbor, Michigan, 48109, United States

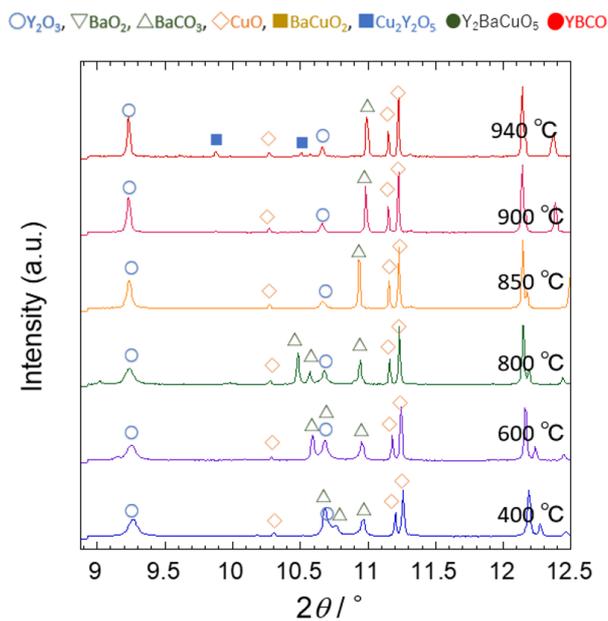

**Figure S1. XRD patterns of the Y$_2$O$_3$-BaCO$_3$-CuO mixture at 400, 600, 800, 850, 900, 940 °C upon heating.** The sample was heated in air at a rate of 30 °C /min. λ = 0.496197 Å.

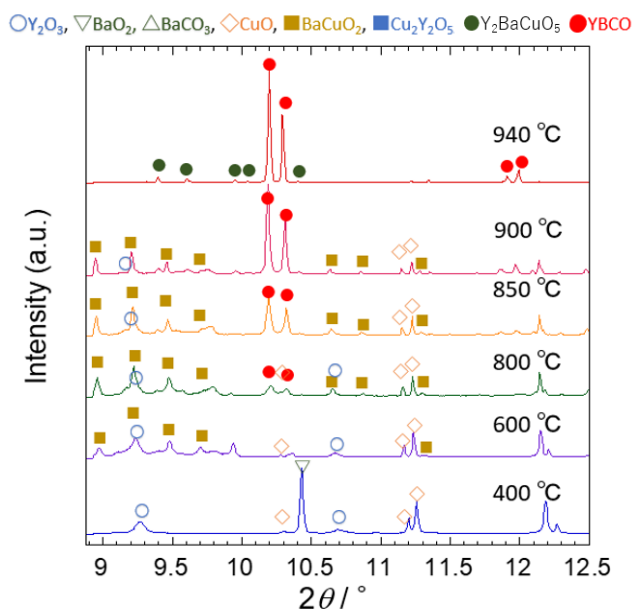

**Figure S2. XRD patterns of the Y$_2$O$_3$-BaO$_2$-CuO mixture at 400, 600, 800, 850, 900, 940 °C upon heating.** The sample was heated in air at a rate of 30 °C /min. λ = 0.496197 Å.

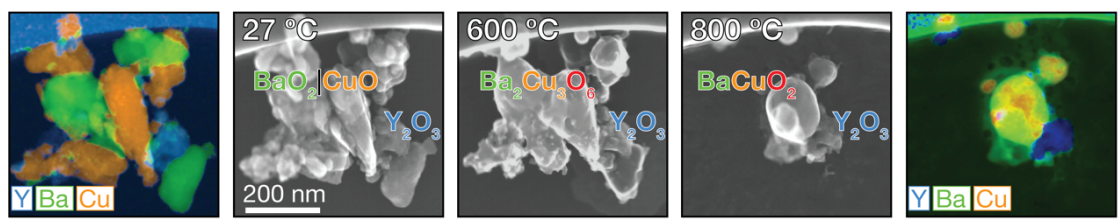

**Figure S3. *In situ* SEM and EDX for the reaction of 0.5 Y$_2$O$_3$ + 2 BaO$_2$ + 3 CuO, heated from 27 °C to 800 °C at 30 °C/min.** The EDX map on the far left was taken before heating and the one on the far right taken after cooling down from 800 °C to room temperature. This provides further confirmation that the only reactive interface among the initial precursors is BaO$_2$|CuO with Y$_2$O$_3$ remaining inert throughout this experiment.

**Table S1. Synthesis recipes extracted for YBCO-related phases**. Target = synthesis target; precursors = synthesis precursors; T = maximum temperature during synthesis; t = total time elapsed during heating operations; tag = Y-Ba-Cu-O if target elements are only these four elements or Y-Ba-Cu-O-* if these elements are present along with others; doi = digital object identifier for paper reporting synthesis. Note that "None" appears for T and t when synthesis operations were not successfully extracted. Recipes using $BaO_2$ as the Ba source are highlighted in yellow.

| target | precursors | T (°C) | t (hr) | tag | doi |
|---|---|---|---|---|---|
| $YBa_2Cu_3O_{7-x}$ | $BaCO_3$; $CuO$; $Y_2O_3$ | 980 | 25 | Y-Ba-Cu-O | 10.1016/j.apcata.2006.03.034 |
| $YBa_2Cu_3O_{7-x}$ | $BaCO_3$; $CuO$; $Y_2O_3$ | None | None | Y-Ba-Cu-O | 10.1016/S0167-577X(99)00202-5 |
| $Y_{1.5}Ba_2Cu_3O_x$ | $Y_2O_3$; $YBa_2Cu_3O_7$ | None | None | Y-Ba-Cu-O | 10.1016/j.jcrysgro.2012.04.029 |
| $YBa_2Cu_3O_{7-x}$ | $BaCO_3$; $CuO$; $Y_2O_3$ | 938 | 24 | Y-Ba-Cu-O | 10.1016/s0167-577x(01)00562-6 |
| $Y_2BaCuO_5$ | $BaCO_3$; $CuO$; $Y_2O_3$ | 1025 | 77 | Y-Ba-Cu-O | 10.1016/s0167-577x(02)00433-0 |
| $Y_2BaCuO_5$ | $BaCO_3$; $CuO$; $Y_2O_3$ | 1025 | 77 | Y-Ba-Cu-O | 10.1016/s0167-577x(02)00433-0 |
| YBaCuO | $BaCO_3$; $CuO$; $Y_2O_3$ | 950 | 20 | Y-Ba-Cu-O | 10.1016/s0038-1098(02)00714-7 |
| $YBa_2Cu_3O_{7-x}$ | $BaCO_3$; $CuO$; $Y_2O_3$ | 930 | 24 | Y-Ba-Cu-O | 10.1016/s0921-5107(97)00213-4 |
| $YBa_2Cu_3O_{7-x}$ | $Ba(CH_3COO)_2$; $Cu(CH_3COO)_2 \cdot H_2O$; $Y(NO_3)_3 \cdot 6H_2O$ | None | None | Y-Ba-Cu-O | 10.1016/s0925-8388(99)00076-6 |
| $YBa_2Cu_3O_{7-x}$ | $BaCO_3$; $CuO$; $Y_2O_3$ | None | None | Y-Ba-Cu-O | 10.1016/s0925-8388(99)00076-6 |
| $YBa_2Cu_3O_{7-x}$ | $BaCO_3$; $CuO$; $Y_2O_3$ | 920 | 12 | Y-Ba-Cu-O | 10.1016/s0167-577x(02)00795-4 |
| $YBa_2Cu_3O_{7-x}$ | $BaCO_3$; $CuO$; $Y_2O_3$ | 1000 | 34 | Y-Ba-Cu-O | 10.1016/j.elspec.2014.02.006 |
| $YBa_2Cu_3O_7$ | $BaCO_3$; $CuO$; $Y_2O_3$ | None | None | Y-Ba-Cu-O | 10.1016/j.radmeas.2004.01.005 |
| $YBa_2Cu_4O_8$ | $Ba(CH_3COO)_2$; $Cu(CH_3COO)_2 \cdot H_2O$; $Y_2O_3$ | 820 | 50 | Y-Ba-Cu-O | 10.1016/s0955-2219(00)00206-5 |

| Material | Precursors | Temp | Time | System | DOI |
|---|---|---|---|---|---|
| YBaCuO | $BaCO_3$; $CuO$; $Y_2O_3$ | 950 | 20 | Y-Ba-Cu-O | 10.1016/S0038-1098(02)00714-7 |
| $YBa_2Cu_3O_{7-x}$ | $BaCO_3$; $CuO$; $Y_2O_3$ | 180 | 25 | Y-Ba-Cu-O | 10.1016/j.eurpolymj.2008.10.020 |
| $YBa_2Cu_3O_{7-x}$ | $BaCO_3$; $CuO$; $Y_2O_3$ | 1060 | 28 | Y-Ba-Cu-O | 10.1016/s0040-6090(99)00717-8 |
| $YBa_2Cu_3O_{7-x}$ | $BaCO_3$; $CuO$; $Y_2O_3$ | 940 | 78 | Y-Ba-Cu-O | 10.1016/s0925-4005(99)00089-1 |
| $YBa_2Cu_3O_{7-x}$ | $Ba(NO_3)_2$; $Cu(NO_3)_2 \cdot 3H_2O$; $Y(NO_3)_3 \cdot 6H_2O$; $NH_3$ | None | None | Y-Ba-Cu-O | 10.1039/c2cp23046a |
| $YBa_2Cu_3O_{7-x}$ | $BaCO_3$; $CuO$; $Y_2O_3$ | 1060 | 28 | Y-Ba-Cu-O | 10.1016/S0040-6090(99)00717-8 |
| $YBa_2Cu_3O_{7-x}$ | $BaCO_3$; $CuO$; $Y_2O_3$ | None | None | Y-Ba-Cu-O | 10.1016/s0167-577x(99)00202-5 |
| Y $Ba_2Cu_3O_{7-x}$ | $BaCO_3$; $CuO$; $Y_2O_3$ | None | None | Y-Ba-Cu-O | 10.1016/j.ssc.2008.11.010 |
| $YBa_2Cu_3O_y$ | $BaO_2$; $CuO$; $Y_2O_3$ | None | None | Y-Ba-Cu-O | 10.1016/j.physc.2004.01.088 |
| $Y_2BaCuO_5$ | $BaO_2$; $CuO$; $Y_2O_3$ | None | None | Y-Ba-Cu-O | 10.1016/j.physc.2004.01.088 |
| $Y_{1.8}Ba_{2.4}Cu_{3.4}O_x$ +0.5 $CeO_2$ + 0.7 w% $Sm_2O_3$ | $BaCO_3$; $CeO_2$; $CuO$; $Sm_2O_3$; $Y_2O_3$ | None | None | Y-Ba-Cu-O | 10.1016/j.jeurceramsoc.2018.01.026 |
| $YBa_2Cu_3O_{7-x}$ | $BaCuO_2$; $CuO$; $Y_2BaCuO_5$ | None | None | Y-Ba-Cu-O | 10.1007/s10854-007-9468-1 |
| $YBa_2Cu_3O_{7-x}$ | $BaCO_3$; $CuO$; $Y_2O_3$ | None | None | Y-Ba-Cu-O | 10.1016/j.physc.2004.03.216 |
| $YBa_2Cu_3O_{7-x}$ | $BaCO_3$; $CuO$; $Y_2O_3$ | 950 | 6 | Y-Ba-Cu-O | 10.1016/j.physc.2004.03.240 |
| $YBa_2Cu_3O_{7-x}$ | $BaCO_3$; $CuO$; $Y_2O_3$ | 940 | 24 | Y-Ba-Cu-O | 10.1016/j.physc.2004.11.003 |
| $YBa_2Cu_3O_{7-x}$ | $BaO_2$; $CuO$; $Y_2O_3$ | 940 | 10 | Y-Ba-Cu-O | 10.1016/j.physc.2010.11.005 |
| $YBa_2Cu_3O_{7-x}$ | $Ba(NO_3)_2$; $Cu(NO_3)_2 \cdot 3H_2O$; $Y(NO_3)_3 \cdot 6H_2O$; $NH_3$ | None | None | Y-Ba-Cu-O | 10.1039/C2CP23046A |
| $Y_2BaCuO_5$ | $Ba(NO_3)_2$; $Cu(NO_3)_2 + 3H_2O$; $Y(NO_3)_3 \cdot 6H_2O$; $NH_3$ | None | None | Y-Ba-Cu-O | 10.1016/j.physc.2013.04.064 |
| $YBa_2Cu_3O_7$ | $Ba(CH_3COO)_2$; $Cu(CH_3COO)_2$; $Y(OH)_3$ | None | None | Y-Ba-Cu-O | 10.1016/j.physc.2015.02.003 |

| Compound | Precursors | Temp (°C) | Time (h) | System | DOI |
|---|---|---|---|---|---|
| YBa$_2$Cu$_3$O$_7$ | Ba(CH$_3$COO)$_2$; Cu(CH$_3$COO)$_2$; Y(CH$_3$COO)$_3$ | None | None | Y-Ba-Cu-O | 10.1016/j.physc.2016.04.004 |
| YBa$_2$Cu$_3$O$_7$ | Ba(NO$_3$)$_2$; Cu(NO$_3$)$_2$·3H$_2$O; Y(NO$_3$)$_3$·6H$_2$O | 930 | 43 | Y-Ba-Cu-O | 10.1016/j.physc.2018.02.010 |
| Y$_3$Ba$_5$Cu$_8$O$_{19}$ | Ba(NO$_3$)$_2$; Cu(NO$_3$)$_2$·3H$_2$O; Y(NO$_3$)$_3$+6H$_2$O | 790 | 14 | Y-Ba-Cu-O | 10.1016/j.physc.2018.02.050 |
| YBa$_2$Cu$_4$O$_8$ | Ba(CH$_3$COO)$_2$; Cu(CH$_3$COO)$_2$·H$_2$O; Y$_2$O$_3$ | 800 | 70 | Y-Ba-Cu-O | 10.1016/s0040-6031(99)00285-3 |
| YBa$_2$Cu$_3$O$_7$ | BaCO$_3$; CuO; Y$_2$O$_3$ | 800 | 20 | Y-Ba-Cu-O | 10.1016/j.physc.2005.09.005 |
| YBa$_2$Cu$_3$O$_{7-x}$ | BaCO$_3$; CuO; Y$_2$O$_3$ | 980 | 21 | Y-Ba-Cu-O | 10.1016/j.memsci.2003.12.011 |
| YBa$_2$Cu$_3$O$_{6.71}$ | BaO; CuO; Y$_2$O$_3$ | 950 | 30 | Y-Ba-Cu-O | 10.1016/j.physc.2006.03.088 |
| YBa$_2$Cu$_3$O$_{7-x}$ | BaO$_2$; CuO; Y$_2$O$_3$ | 1100 | 36 | Y-Ba-Cu-O | 10.1016/j.jpcs.2013.04.025 |
| YBa$_2$Cu3Oy | BaCO$_3$; CuO; Y$_2$O$_3$ | 950 | 32 | Y-Ba-Cu-O | 10.1016/j.physc.2007.01.033 |
| YBa$_2$Cu$_3$O$_{7-x}$ | BaCO$_3$; CuO; Y$_2$O$_3$ | 950 | 32 | Y-Ba-Cu-O | 10.1016/j.physc.2007.03.108 |
| YBa$_2$Cu$_3$O$_{7-x}$ | BaCO$_3$; CuO; Y$_2$O$_3$ | None | None | Y-Ba-Cu-O | 10.1016/j.physc.2007.04.234 |
| YBa$_2$Cu$_4$O$_8$ | BaCO$_3$; CuO; Y$_2$O$_3$ | 935 | 110 | Y-Ba-Cu-O | 10.1103/PhysRevB.70.144515 |
| YBa$_2$Cu$_3$O$_{7-d}$ | BaCO$_3$; CuO; Y$_2$O$_3$ | 930 | 48 | Y-Ba-Cu-O | 10.1021/cm020747j |
| YBa$_2$Cu$_3$O$_{7-x}$ | BaCuO$_2$; CuO; Y$_2$O$_3$ | None | None | Y-Ba-Cu-O | 10.1016/j.physc.2007.05.001 |
| YBa$_2$Cu$_3$O$_{7-x}$ | BaCO$_3$; CuO; Y$_2$O$_3$ | None | None | Y-Ba-Cu-O | 10.1016/j.jqsrt.2004.09.023 |
| YBa$_2$Cu$_3$O$_{7-x}$ | BaCO$_3$; CuO; Y$_2$O$_3$ | 1045 | 200 | Y-Ba-Cu-O | 10.1016/j.physc.2007.07.010 |
| Y$_2$BaCuO$_{7-x}$ | BaCO$_3$; CuO; Y$_2$O$_3$ | 1045 | 200 | Y-Ba-Cu-O | 10.1016/j.physc.2007.07.010 |
| YBa$_2$Cu$_4$O$_8$ | Ba(CH$_3$COO)$_2$; Cu(CH$_3$COO)$_2$·H$_2$O; Y$_2$O$_3$ | None | None | Y-Ba-Cu-O | 10.1016/j.chemphys.2006.04.007 |

| Compound | Precursors | Temp (°C) | Time (h) | System | DOI |
|---|---|---|---|---|---|
| YBa$_2$Cu$_3$O$_x$ | BaCO$_3$; CuO; Y$_2$O$_3$ | None | None | Y-Ba-Cu-O | 10.1016/S0921-4534(00)01520-3 |
| Y$_2$BaCuO$_5$ | BaCO$_3$; CuO; Y$_2$O$_3$ | 1050 | 12 | Y-Ba-Cu-O | 10.1016/s0925-8388(98)00427-7 |
| YBa$_2$Cu$_3$O$_7$-x | BaCO$_3$; CuO; Y$_2$O$_3$ | None | None | Y-Ba-Cu-O | 10.1016/j.physc.2008.01.007 |
| Y$_2$BaCuO$_5$ | BaCO$_3$; CuO; Y$_2$O$_3$ | None | None | Y-Ba-Cu-O | 10.1016/j.physc.2008.01.007 |
| YBa$_2$Cu$_3$O$_x$ | BaCO$_3$; CuO; Y$_2$O$_3$ | 900 | 8 | Y-Ba-Cu-O | 10.1016/s0925-8388(98)00664-1 |
| YBa$_2$Cu$_3$O$_x$ | BaCO$_3$; CuO; Y$_2$O$_3$ | 950 | 32 | Y-Ba-Cu-O | 10.1016/j.physc.2008.12.002 |
| YBa$_2$Cu$_3$O$_7$ | BaCO$_3$; CuO; Y$_2$O$_3$ | 900 | 72 | Y-Ba-Cu-O | 10.1021/ja9706920 |
| YBa$_2$Cu$_3$O$_x$ | BaCO$_3$; CuO; Y$_2$O$_3$ | 945 | 16 | Y-Ba-Cu-O | 10.1016/s0925-8388(99)00115-2 |
| Y$_2$BaCuO$_5$ | BaO; CuO; Y$_2$O$_3$ | 880 | 24 | Y-Ba-Cu-O | 10.1016/j.physc.2009.05.019 |
| YBa$_{2-x}$Na$_x$Cu$_3$O$_y$+40mol%Y$_2$BaCuO$_5$ | BaCO$_3$; CuO; Na$_2$C$_2$O$_4$; Y$_2$O$_3$ | 1050 | 150 | Y-Ba-Cu-O | 10.1016/S0921-4534(01)00150-2 |
| YBa$_2$Cu$_3$O$_7$-y | BaCO$_3$; CuO; Y$_2$O$_3$ | 950 | 30 | Y-Ba-Cu-O | 10.1016/j.physc.2009.05.106 |
| YBa$_2$Cu$_3$O$_{7-x}$ | BaCO$_3$; CuO; Y$_2$O$_3$ | 930 | 72 | Y-Ba-Cu-O | 10.1016/j.cryogenics.2015.05.011 |
| Y$_3$Ba$_5$Cu$_8$O$_{18}$ | BaCO$_3$; CuO; Y$_2$O$_3$ | 840 | 12 | Y-Ba-Cu-O | 10.1016/j.physc.2009.09.003 |
| YBa$_2$Cu$_3$O$_{7-x}$ | BaCO$_3$; CuO; Y$_2$O$_3$ | 950 | 8 | Y-Ba-Cu-O | 10.1016/j.physc.2009.11.034 |
| YBa$_2$Cu$_3$O$_{7-x}$ | BaCO$_3$; CuO; Y$_2$O$_3$ | 925 | 32 | Y-Ba-Cu-O | 10.1016/j.jmmm.2010.04.002 |
| Y$_2$Ba$_5$Cu$_7$O$_x$ | BaCO$_3$; CuO; Y$_2$O$_3$ | 850 | 48 | Y-Ba-Cu-O | 10.1016/j.ssc.2016.02.017 |
| Y$_2$BaCuO$_5$ | BaCO$_3$; CuO; Y$_2$O$_3$ | 900 | 60 | Y-Ba-Cu-O | 10.1016/S0921-4534(01)00624-4 |
| Y$_2$BaCuO$_5$ | BaCO$_3$; CuO; YBa$_2$Cu$_3$O$_{7-x}$ | None | None | Y-Ba-Cu-O | 10.1016/j.physc.2010.05.012 |
| YBa$_2$Cu$_3$O$_{7-x}$ | BaCO$_3$; CuO; Y$_2$O$_3$ | 950 | 16 | Y-Ba-Cu-O | 10.1016/S0921-4534(01)00831-0 |
| YBa$_2$Cu$_4$O$_8$ | Ba(CH$_3$COO)$_2$; Cu(CH$_3$COO)$_2$·H$_2$O; Y$_2$O$_3$ | None | None | Y-Ba-Cu-O | 10.1016/S0924-2031(01)00157-6 |
| YBa$_2$Cu$_3$O$_y$ | BaO; CuO; Y$_2$O$_3$ | 910 | 12 | Y-Ba-Cu-O | 10.1016/j.physc.2010.05.236 |

| Compound | Precursors | Temp (°C) | Time (h) | System | DOI |
|---|---|---|---|---|---|
| $Y_2BaCuO_5$ | BaO; CuO; $Y_2O_3$ | 910 | 12 | Y-Ba-Cu-O | 10.1016/j.physc.2010.05.236 |
| $Y_2BaCuO_5$ | $BaCO_3$; CuO; $Y_2O_3$ | None | None | Y-Ba-Cu-O | 10.1016/j.jcrysgro.2005.01.094 |
| $YBa_2Cu_3O_y$ | $Ba(NO_3)_2$; $Cu(NO_3)_2 \cdot 3H_2O$; $Y_2O_3$ | None | None | Y-Ba-Cu-O | 10.1016/j.mseb.2003.11.015 |
| $YBa_2Cu_3O_7$ | $BaCO_3$; CuO; $Y_2O_3$ | 940 | 144 | Y-Ba-Cu-O | 10.1039/c4ta06767c |
| $Y_2BaCuO_5$ | $BaCuO_2$; $Y_2O_3$ | 820 | 20 | Y-Ba-Cu-O | 10.1016/S0921-4534(01)00968-6 |
| $YBa_2Cu_3O_{7-x}$ | $BaO_2$; CuO; $Y_2O_3$ | 940 | 10 | Y-Ba-Cu-O | 10.1016/j.physc.2010.11.005 |
| $YBa_2Cu_3O_y$ | $BaCO_3$; CuO; $Y_2O_3$ | 950 | 32 | Y-Ba-Cu-O | 10.1016/j.physc.2010.12.012 |
| $YBa_2Cu_3O_{7-x}$ | $BaCO_3$; CuO; $Y_2O_3$ | 950 | 10 | Y-Ba-Cu-O | 10.1016/S0921-4534(02)01318-7 |
| $YBa_2Cu3O_y$ | $BaCO_3$; CuO; $Y_2O_3$ | 950 | 20 | Y-Ba-Cu-O | 10.1016/j.physc.2011.10.003 |
| $YBa_2Cu_3O_{7-x}$ | $BaCO_3$; CuO; $Y_2O_3$ | None | None | Y-Ba-Cu-O | 10.1016/j.physc.2012.05.012 |
| $YBa_2Cu_3O_{7-x}$ | $BaCO_3$; CuO; $Y_2O_3$ | 938 | 24 | Y-Ba-Cu-O | 10.1016/S0167-577X(01)00562-6 |
| $YBa_2Cu_3O_{7-x}$ | BaO; CuO; $Y_2O_3$ | None | None | Y-Ba-Cu-O | 10.1016/j.solidstatesciences.2005.07.002 |
| $Y_2BaCuO_5$ | $BaCO_3$; CuO; $Y_2O_3$ | 1025 | 77 | Y-Ba-Cu-O | 10.1016/S0167-577X(02)00433-0 |
| $Y_2BaCuO_5$ | $BaCO_3$; CuO; $Y_2O_3$ | 1025 | 77 | Y-Ba-Cu-O | 10.1016/S0167-577X(02)00433-0 |
| $YBa_2Cu_3O_{7-x}$ | BaO; CuO; $Y_2O_3$ | None | None | Y-Ba-Cu-O | 10.1016/j.jssc.2010.01.006 |
| $Y_{1.5}Ba_2Cu_3O_{7-x}$ | $Y_2O_3$; $YBa_2Cu_3O_7$ | None | None | Y-Ba-Cu-O | 10.1016/j.physc.2013.04.028 |
| $Y_{1.5}Ba_2Cu_3O_{7-x}$ | $Y_2O_3$; $YBa_2Cu_3O_7$ | None | None | Y-Ba-Cu-O | 10.1016/j.physc.2013.04.084 |
| $YBa_2Cu_3O_y$ | $BaCO_3$; CuO; $Y_2O_3$ | 950 | 20 | Y-Ba-Cu-O | 10.1016/j.physc.2013.12.006 |
| $YBa_2Cu_3O_y$ | $BaCO_3$; CuO; $Y_2O_3$ | 900 | 94 | Y-Ba-Cu-O | 10.1016/S0921-4534(02)02058-0 |
| $Y_3Ba_5Cu_8O_{18\pm x}$ | $BaCO_3$; CuO; $Y_2O_3$ | 950 | 60 | Y-Ba-Cu-O | 10.1007/s00339-017-1547-4 |
| $YBa_2Cu_3O_{7-x}$ | $BaCO_3$; CuO; $Y_2O_3$ | 950 | 60 | Y-Ba-Cu-O | 10.1007/s00339-017-1547-4 |
| $Y_2BaCuO_5$ | $BaCO_3$; CuO; $Y_2O_3$ | 900 | 94 | Y-Ba-Cu-O | 10.1016/S0921-4534(02)02058-0 |

| Material | Precursors | Temp (°C) | Time (h) | System | DOI |
|---|---|---|---|---|---|
| $YBa_2Cu_3O_{7-x}$ | $BaCO_3$; $CuO$; $Y_2O_3$ | 980 | 34 | Y-Ba-Cu-O | 10.1016/j.ssc.2004.04.044 |
| $YBa_2Cu_3O_{7-x}$ | $Ag_2O$; $BaO$; $CuO$; $Y_2O_3$ | 950 | 74 | Y-Ba-Cu-O | 10.1016/j.ssc.2004.05.015 |
| $Y_2BaCuO_5$ | $BaCO_3$; $CuO$; $Y_2O_3$ | 900 | 24 | Y-Ba-Cu-O | 10.1016/j.physc.2014.05.009 |
| $Y_{1.6}Ba_{2.3}Cu_{3.3}O_y$ | $BaCO_3$; $CuO$; $Y_2O_3$ | 1400 | 26 | Y-Ba-Cu-O | 10.1016/S0921-4534(02)02539-x |
| $YBa_2Cu_3O_{7-x}$ | $BaCu3$; $CuO$; $Y_2O_3$ | 920 | 12 | Y-Ba-Cu-O | 10.1016/S0167-577X(02)00795-4 |
| $YBa_2Cu_3O_y$ | $BaCO_3$; $CuO$; $Y_2O_3$ | 925 | 74 | Y-Ba-Cu-O | 10.1016/j.physc.2016.11.003 |
| $YBa_2Cu_3O_{6+x}$ | $BaCO_3$; $CuO$; $Y_2O_3$ | None | None | Y-Ba-Cu-O | 10.1103/PhysRevB.93.054523 |
| $Y_3Ba_5Cu_8O_{19}$ | $BaCO_3$; $CuO$; $Y_2O_3$ | 840 | 24 | Y-Ba-Cu-O | 10.1016/j.physc.2018.02.050 |
| $YBa_2Cu_3O_{7-x}$ | $BaCO_3$; $CuO$; $Y_2O_3$ | None | None | Y-Ba-Cu-O | 10.1111/j.1551-2916.2008.02900.x |
| Y3Ba5Cu8O18 | $Ba(NO_3)_2$; $CuO$; $Y_2O_3$ | 900 | 72 | Y-Ba-Cu-O | 10.1016/j.solidstatesciences.2011.08.024 |
| $YBa_2Cu_3O_{7-x}$ | $BaCO_3$; $CuO$; $Y_2O_3$ | 945 | 24 | Y-Ba-Cu-O | 10.1007/s10854-013-1212-4 |
| $YBa_2Cu_3O_7$ | $BaCO_3$; $CuO$; $Y_2O_3$ | 900 | 72 | Y-Ba-Cu-O | 10.1021/ja9706920 |
| $YBa_2Cu_3O_{7-x}$ | $BaCO_3$; $CuO$; $Y_2O_3$ | 980 | 21 | Y-Ba-Cu-O | 10.1016/j.ssi.2004.10.003 |
| $YBa_2Cu_4O_8$ | $Ba(CH_3COO)_2$; $Cu(CH_3COO)_2 \cdot H_2O$; $Y_2O_3$ | None | None | Y-Ba-Cu-O | 10.1016/s0924-2031(01)00157-6 |
| $Y_{1-x}Pr_xBa_2Cu_3O_{7-x}$ | $BaCO_3$; $CuO$; $Pr_6O_{11}$; $Y_2O_3$ | 935 | 36 | Y-Ba-Cu-O-* | 10.1016/s0167-577x(01)00577-8 |
| $Y_2Ba(Cu_{1-x}Mg_x)O_5$ | $BaCO_3$; $CuO$; $MgO$; $Y_2O_3$ | 1000 | 12 | Y-Ba-Cu-O-* | 10.1016/s0955-2219(03)00548-x |
| $YBa_{2-x}La_xCu_3O_y$ | $BaCO_3$; $CuO$; $La_2O_3$; $Y_2O_3$ | 920 | 42 | Y-Ba-Cu-O-* | 10.1016/S0038-1098(00)00360-4 |
| $(La_{1-x}Yx)_2Ba_2CaCu_5O_z$ | $BaCO_3$; $CaCO_3$; $CuO$; $La_2O_3$; $Y_2O_3$ | 900 | 48 | Y-Ba-Cu-O-* | 10.1016/j.ssc.2006.03.035 |
| $Y_{1-y}Yb_yBa_2Cu_3O_x$ | $BaCO_3$; $CuO$; $Y_2O_3$; $Yb_2O_3$ | None | None | Y-Ba-Cu-O-* | 10.1103/PhysRevB.79.054519 |
| $Y_2Ba(Cu_{1-x}Ni_x)O_5$ | $BaCO_3$; $CuO$; $NiO$; $Y_2O_3$ | 1300 | 32 | Y-Ba-Cu-O-* | 10.1016/S0921-5107(00)00566-3 |
| $Ba(Zr_{0.84}Y_{0.15}Cu_{0.01})O_{3-x}$ | $BaCO_3$; $CuO$; $Y_2O_3$; $ZrO_2$ | 1300 | 2 | Y-Ba-Cu-O-* | 10.1007/s10008-013-2187-z |

| Compound | Precursors | Temp (°C) | Time (h) | Class | DOI |
|---|---|---|---|---|---|
| Y(Ba$_{1-x}$Sr$_x$)$_2$Cu$_3$O$_{7-x}$ | BaCO$_3$; CuO; SrCO$_3$; Y$_2$O$_3$ | 930 | 24 | Y-Ba-Cu-O-* | 10.1016/j.ssc.2006.07.026 |
| YBa$_2$(Cu$_{1-x}$Ni$_x$)$_3$O$_{7-x}$ | BaCO$_3$; CuO; Ni$_2$O$_3$; Y$_2$O$_3$ | None | None | Y-Ba-Cu-O-* | 10.1016/S0038-1098(01)00490-2 |
| YBaCuFeO$_5$ | BaCO$_3$; CuO; Fe$_2$O$_3$; Y$_2$O$_3$ | 1150 | 72 | Y-Ba-Cu-O-* | 10.1016/j.jcrysgro.2014.12.020 |
| (La$_{1-x}$Y$_x$)$_2$Ba$_2$CaCu$_5$O$_z$ | BaCO$_3$; CaCO$_3$; CuO; La$_2$O$_3$; Y$_2$O$_3$ | 900 | 48 | Y-Ba-Cu-O-* | 10.1016/j.ssc.2006.09.008 |
| Y$_{1-x}$Nd$_x$Ba$_2$Cu$_3$O$_{7-x}$ | BaCO$_3$; CuO; Nd$_2$O$_3$; Y$_2$O$_3$ | 930 | 46 | Y-Ba-Cu-O-* | 10.1016/j.jmatprotec.2007.12.078 |
| Y$_{1-x}$Ca$_x$Ba$_2$Cu$_{2.85}$Re$_{0.15}$O$_z$ | BaCO$_3$; CaO; CuO; ReO$_3$; Y$_2$O$_3$ | None | None | Y-Ba-Cu-O-* | 10.1016/s0038-1098(99)00085-x |
| YBaCuFeO$_5$ | BaCO$_3$; CuO; Fe$_2$O$_3$; Y$_2$O$_3$ | 1150 | 100 | Y-Ba-Cu-O-* | 10.1038/ncomms13758 |
| YBa$_2$(Cu$_{1-x}$Mn$_x$)$_4$O$_8$ | Ba(CH$_3$COO)$_2$; Cu(CH$_3$COO)$_2$·H$_2$O; Mn(CH$_3$COO)$_2$; Y$_2$O$_3$ | 820 | 50 | Y-Ba-Cu-O-* | 10.1016/s0955-2219(00)00206-5 |
| Y$_{0.7}$Ca$_{0.3}$Ba$_2$Cu$_3$O$_y$F$_x$ | BaCO$_3$; CaCO$_3$; CaF$_2$; CuO; Y$_2$O$_3$ | 920 | 84 | Y-Ba-Cu-O-* | 10.1016/j.jmmm.2003.11.105 |
| Y$_{0.5}$Nd$_{0.5}$Ba$_2$Cu$_3$O$_x$ | BaCO$_3$; CuO; Nd$_2$O$_3$; Y$_2$O$_3$ | 900 | 48 | Y-Ba-Cu-O-* | 10.1016/S0022-0248(99)00391-7 |
| (La$_{2-x}$Y$_x$)Ba$_2$(Ca$_y$Cu$_{4+y}$)O$_z$ | BaCO$_3$; CaCO$_3$; CuO; La$_2$O$_3$; Y$_2$O$_3$ | 950 | 84 | Y-Ba-Cu-O-* | 10.1016/s0167-577x(98)00067-6 |
| YBa$_{2-x}$La$_x$Cu$_3$O$_{7-x}$ | BaCO$_3$; CuO; La$_2$O$_3$; Y$_2$O$_3$ | 1203 | 72 | Y-Ba-Cu-O-* | 10.1016/s1293-2558(03)00187-0 |
| YBa$_{2-x}$La$_x$Cu$_3$O$_y$ | BaCO$_3$; CuO; La$_2$O$_3$; Y$_2$O$_3$ | 950 | 1000 | Y-Ba-Cu-O-* | 10.1016/j.physc.2003.09.002 |
| YBa$_{2-x}$Na$_x$Cu$_3$O$_y$+40 | BaCO$_3$; CuO; Na$_2$C$_2$O$_4$; Y$_2$O$_3$ | 1040 | 174 | Y-Ba-Cu-O-* | 10.1016/s0167-577x(99)00178-0 |
| Cu$_{1-0.75x}$(Sr$_{2x}$Ba$_{2-2x}$)(Ca$_{0.5x}$Y$_{1-0.5x}$)Cu$_2$O$_y$ | BaCO$_3$; CaCO$_3$; CuO; SrCO$_3$; Y$_2$O$_3$ | 970 | 15 | Y-Ba-Cu-O-* | 10.1016/s0022-3697(01)00117-2 |
| YBa$_2$Cu$_3$F$_{0.4}$O$_x$ | YBa$_2$Cu$_3$F$_4$O$_x$; YBa$_2$Cu$_3$O$_x$ | 900 | 8 | Y-Ba-Cu-O-* | 10.1016/S0924-0136(99)00474-4 |
| Y$_{1-x}$La$_x$Ba$_2$Cu$_3$O$_{7-x}$ | BaCO$_3$; CuO; La$_2$O$_3$; Y$_2$O$_3$ | 970 | 41 | Y-Ba-Cu-O-* | 10.1016/j.mseb.2006.12.007 |
| Y$_{1-y}$CaYBa$_2$Cu$_3$O$_{7-x}$ | BaCO$_3$; CaCO$_3$; CuO; Y$_2$O$_3$ | 970 | 41 | Y-Ba-Cu-O-* | 10.1016/j.mseb.2006.12.007 |
| Y$_{1-x}$Ca$_x$BaCuFeO$_{5+x}$ | BaCO$_3$; CaCO$_3$; CuO; Fe$_2$O$_3$; Y$_2$O$_3$ | None | None | Y-Ba-Cu-O-* | 10.1016/j.solidstatesciences.2011.10.021 |
| Y$_{1-x}$Ca$_x$Ba$_2$Cu$_3$O$_z$ | BaCO$_3$; CaCO$_3$; CuO; Y$_2$O$_3$ | 950 | 48 | Y-Ba-Cu-O-* | 10.1016/j.physc.2004.01.002 |
| TlBa$_2$Y$_{1-x}$Ca$_x$Cu$_2$O$_{7+x}$ | BaO$_2$; CaO; CuO; Tl$_2$O$_3$; Y$_2$O$_3$ | None | None | Y-Ba-Cu-O-* | 10.1016/s0022-3697(02)00087-2 |

| Compound | Precursors | Temperature | Time | Class | DOI |
|---|---|---|---|---|---|
| $Y_2Ba_4CuWO_{10.8}$ | $BaCO_3$; $CuO$; $WO_3$; $Y_2O_3$ | None | None | Y-Ba-Cu-O-* | 10.1016/j.jeurceramsoc.2018.01.026 |
| $Y_2Ba_4CuWO_x$ | $BaCO_3$; $CuO$; $WO_3$; $Y_2O_3$ | None | None | Y-Ba-Cu-O-* | 10.1016/j.jeurceramsoc.2018.01.026 |
| $Ba(Zr_{0.84}Y_{0.15}Cu_{0.01})O_{3-x}$ | $BaCO_3$; $CuO$; $Y_2O_3$; $ZrO_2$ | 1500 | 42 | Y-Ba-Cu-O-* | 10.1016/j.jpowsour.2016.09.129 |
| $Y_{1-x}Ca_xBa_2Cu_3O_y$ | $BaCO_3$; $CaCO_3$; $CuO$; $Y_2O_3$ | None | None | Y-Ba-Cu-O-* | 10.1016/s0254-0584(01)00545-4 |
| $YBa_2Cu_{3-x}Gd_xO_{7-x}$ | $BaCO_3$; $CuO$; $Gd_2O_3$; $Y_2O_3$ | 930 | 24 | Y-Ba-Cu-O-* | 10.1016/j.physc.2004.10.008 |
| $(Y_{0.74}Ca_{0.26})Ba_2Cu_3O_{7-x}$ | $BaCO_3$; $CaCO_3$; $CuO$; $Y_2O_3$ | 980 | 102 | Y-Ba-Cu-O-* | 10.1016/j.jpcs.2010.10.079 |
| $(Y_{0.84}La_{0.16})(Ba_{1.74}La_{0.26})Cu_3O_{7-x}$ | $BaCO_3$; $CuO$; $La_2O_3$; $Y_2O_3$ | 980 | 102 | Y-Ba-Cu-O-* | 10.1016/j.jpcs.2010.10.079 |
| $YBa_2Cu_{3-x}Gd_xO_{7-x}$ | $BaCO_3$; $CuO$; $Gd_2O_3$; $Y_2O_3$ | 930 | 24 | Y-Ba-Cu-O-* | 10.1016/j.physc.2004.10.008 |
| $Fe_{0.5}Cu_{0.5}Ba_2YCu_2O7.41$ | $BaCO_3$; $CuO$; $Fe_2O_3$; $Y_2O_3$ | 930 | 110 | Y-Ba-Cu-O-* | 10.1016/j.physc.2004.11.002 |
| $Y0.8Ca0.2Ba_2Cu_3O_y$ | $BaCO_3$; $CaCO_3$; $CuO$; $Y_2O_3$ | 920 | 96 | Y-Ba-Cu-O-* | 10.1016/j.physc.2004.11.006 |
| $YBa_2(Cu_{1-x}Zn_x)_3O_{7-x}$ | $BaCO_3$; $CuO$; $Y_2O_3$; $ZnO$ | 1050 | 148 | Y-Ba-Cu-O-* | 10.1016/j.physc.2010.01.032 |
| $Y_{1-x}Ca_xBa_2Cu_3O_{7-x}$ | $BaCO_3$; $CaO$; $CuO$; $Y_2O_3$ | 920 | 48 | Y-Ba-Cu-O-* | 10.1016/j.physc.2005.01.002 |
| $(Hg_{0.5}Pb_{0.5})(Sr_{2-x}Ba_x)(Ca_{0.7}Y_{0.3})Cu_2O_{7-d}$ | $BaO_2$; $CaO$; $CuO$; $HgO$; $PbO$; $SrO_2$; $Y_2O_3$ | 970 | 24 | Y-Ba-Cu-O-* | 10.1021/ic9611249 |
| $Y_2Ba_4CuNbO_y$ | $BaCO_3$; $CuO$; $Nb_2O_5$; $Y_2O_3$ | None | None | Y-Ba-Cu-O-* | 10.1016/j.physc.2005.02.060 |
| $Y_{0.92}Ta_{0.08}Ba_2Cu_3O_y$ | $BaCO_3$; $CuO$; $Ta_2O_5$; $Y_2O_3$ | 900 | 36 | Y-Ba-Cu-O-* | 10.1016/j.physc.2005.03.010 |
| $(Cu_{1-x}Co_x)(Ba_{1-y}Sr_y)_2(Y_{1-z}Ca_z)Cu_2O_{7+x}$ | $BaCO_3$; $CaCO_3$; $Co_3O_4$; $CuO$; $SrCO_3$; $Y_2O_3$ | 940 | 48 | Y-Ba-Cu-O-* | 10.1016/j.physc.2005.04.034 |
| $YBaCuCoO_{5+x}$ | $Ba(NO_3)_2$; $Co(NO_3)_2·6H_2O$; $Cu(NO_3)_2·6H_2O$; $Y_2O_3$ | 1000 | 3 | Y-Ba-Cu-O-* | 10.1002/fuce.201400141 |
| $Y_{1+x}Sb_xBa_2Cu_3O_z$ | $BaCO_3$; $CuO$; $Sb_2O_3$; $Y_2O_3$ | 800 | 20 | Y-Ba-Cu-O-* | 10.1016/j.physc.2005.09.005 |
| $YBa_2Cu_{3-x}Ca_xO_{7-y}$ | $BaCO_3$; $CaO$; $CuO$; $Y_2O_3$ | 940 | 72 | Y-Ba-Cu-O-* | 10.1016/j.sna.2012.06.015 |
| $YBa_2Cu_{2.99}Li_{0.01}O_y + 0.4Y_2BaCuO_5$ | $BaCO_3$; $CuO$; $Li_2CO_3$; $Y_2O_3$ | 1035 | 48 | Y-Ba-Cu-O-* | 10.1016/j.physc.2006.02.012 |
| $YBa_{2-x}K_xCu_3O_y$ | $BaCO_3$; $CuO$; $K_2CO_3$; $Y_2O_3$ | 920 | 40 | Y-Ba-Cu-O-* | 10.1016/j.physc.2006.03.093 |

| Compound | Precursors | Temp (°C) | Time (h) | System | DOI |
|---|---|---|---|---|---|
| $Y_{1-x}B_xBa_2Cu_3O_y$ | $B_2O_3$; $Ba_2CO_3$; $CuO$; $Y_2O_3$ | 950 | 28 | Y-Ba-Cu-O-* | 10.1016/j.physc.2006.03.135 |
| $Y_{0.95}Pr_{0.05}Ba_2(Cu_{1-x}Mn_x)_3O_{7-x}$ | $BaCO_3$; $CuO$; $MnO_2$; $Pr_6O_{11}$; $Y_2O_3$ | 915 | 24 | Y-Ba-Cu-O-* | 10.1016/j.physc.2006.08.002 |
| $Cu_{1-0.75x}(Sr_{2x}Ba_{2-2x})(Ca_{0.5x}Y_{1-0.5x})Cu_2O_y$ | $BaCO_3$; $CaCO_3$; $CuO$; $SrCO_3$; $Y_2O_3$ | 970 | 15 | Y-Ba-Cu-O-* | 10.1016/S0022-3697(01)00117-2 |
| $YBa_2(Cu_{1-x}Zn_x)_3O_{6+x}$ | $BaCO_3$; $CuO$; $Y_2O_3$; $ZnO$ | None | None | Y-Ba-Cu-O-* | 10.1016/S0921-4534(00)00118-0 |
| $YBa_2Cu_{3-x}M_xO_y$ | $Al_2O_3$; $BaCO_3$; $CuO$; $Y_2O_3$ | 920 | 12 | Y-Ba-Cu-O-* | 10.1103/PhysRevB.69.224517 |
| $YBa_2Cu_{3-x}M_xO_y$ | $BaCO_3$; $CuO$; $Y_2O_3$; $ZnO$ | 920 | 12 | Y-Ba-Cu-O-* | 10.1103/PhysRevB.69.224517 |
| $Y_{0.38}La_{0.62}(Ba_{0.82}La_{0.18})_2Cu_3O_y$ | $BaCO_3$; $CuO$; $La_2O_3$; $Y_2O_3$ | 980 | 90 | Y-Ba-Cu-O-* | 10.1103/PhysRevB.86.045124 |
| $TlBa_2Y_{1-x}Ca_xCu_2O_{7+x}$ | $BaO_2$; $CaO$; $CuO$; $Tl_2O_3$; $Y_2O_3$ | None | None | Y-Ba-Cu-O-* | 10.1016/S0022-3697(02)00087-2 |
| $HgBa_2(Ca_{1-x}Yx)Cu_2O_y$ | $BaO$; $CaO$; $CuO$; $HgO$; $Y_2O_3$ | 720 | 22 | Y-Ba-Cu-O-* | 10.1016/S0921-4534(00)00205-7 |
| $YBa_2(Cu_{3-x}Scx)\ O_y$ | $BaCO_3$; $CuO$; $Sc_2O_3$; $Y_2O_3$ | 967 | 72 | Y-Ba-Cu-O-* | 10.1016/j.physc.2007.04.043 |
| $Y_{(1-x)}Ce_xBa_2Cu_3O_7$ | $BaCO_3$; $CeO_2$; $CuO$; $Y_2O_3$ | 930 | 160 | Y-Ba-Cu-O-* | 10.1016/j.physc.2007.04.046 |
| $YBa_{2-x}La_xCu_3O_{7-x}$ | $BaCO_3$; $CuO$; $La_2O_3$; $Y_2O_3$ | 1203 | 72 | Y-Ba-Cu-O-* | 10.1016/S1293-2558(03)00187-0 |
| $Y(Ba_{2-x}Sr_x)Cu_3O_{6.95}$ | $BaCO_3$; $CuO$; $SrCO_3$; $Y_2O_3$ | 950 | 60 | Y-Ba-Cu-O-* | 10.1016/S0921-4534(00)00293-8 |
| $Y_{1-x}Ca_xBa_2Cu_3\ O_y$ | $BaCO_3$; $CaCO_3$; $CuO$; $Y_2O_3$ | 750 | 15 | Y-Ba-Cu-O-* | 10.1103/PhysRevB.70.214517 |
| $YBa_{2-x}M_xCu_3O_y$ | $BaCO_3$; $CuO$; $NaNO_3$; $Y_2O_3$ | 950 | 12 | Y-Ba-Cu-O-* | 10.1016/S0921-4534(00)00338-5 |
| $Y_2Ba_4CuMO_x$ | $BaCO_3$; $CuO$; $Y_2O_3$; $ZrO_2$ | None | None | Y-Ba-Cu-O-* | 10.1111/j.1551-2916.2007.01771.x |
| $Y_2Ba_4CuMO_x$ | $BaCO_3$; $CuO$; $Nb_2O_5$; $Y_2O_3$ | None | None | Y-Ba-Cu-O-* | 10.1111/j.1551-2916.2007.01771.x |
| $Y_{1-x}Ca_xBa_2Cu_3O_{7-x}$ | $Ba(NO_3)_2$; $CaCO_3$; $CuO$; $Y_2O_3$ | 450 | 1 | Y-Ba-Cu-O-* | 10.1016/S0921-4534(00)01530-6 |
| $Ba(Zr_{0.84}Y_{015}Cu_{0.01})O_{3-x}$ | $BaCO_3$; $CuO$; $Y_2O_3$; $ZrO_2$ | None | None | Y-Ba-Cu-O-* | 10.1016/j.ceramint.2013.05.081 |
| $YBa_{1.8}La_{0.2}Cu_3O_y$ | $BaCO_3$; $CuO$; $La_2O_3$; $Y_2O_3$ | 950 | 20 | Y-Ba-Cu-O-* | 10.1016/S0921-4534(00)01549-5 |
| $Ba(Nd_xY_{2-x})CuO_5$ | $BaCO_3$; $CuO$; $Nd_2O_3$; $Y_2O_3$ | 980 | 44 | Y-Ba-Cu-O-* | 10.1016/j.jssc.2008.08.002 |
| $YBa_2Cu_{3-x}Zn_xO_{6+x}$ | $CuO$; $Y_2O_3$; $ZnO$; $BaCO_3$ | None | None | Y-Ba-Cu-O-* | 10.1016/s0925-8388(98)00577-5 |

| Compound | Precursors | Temp (°C) | Time (h) | Class | DOI |
|---|---|---|---|---|---|
| $Y_{1-x}Tb_xBa_2Cu_3O_{7-x}$ | $BaCO_3$; $CuO$; $Tb_4O_7$; $Y_2O_3$ | 950 | 60 | Y-Ba-Cu-O-* | 10.1016/j.physc.2008.04.012 |
| $YBa_2Cu_{3-x}Zn_xO_{7-x}$ | $BaCO_3$; $CuO$; $Y_2O_3$; $ZnO$ | 970 | 26 | Y-Ba-Cu-O-* | 10.1016/S0921-4534(00)01648-8 |
| $(La_{1-x}Y_x)_2Ba_2CaCu_5Oz$ | $BaCO_3$; $CaCO_3$; $CuO$; $La_2O_3$; $Y_2O_3$ | 900 | 24 | Y-Ba-Cu-O-* | 10.1111/j.1551-2916.2007.01845.x |
| $YBa_{2-x}Sr_xCu_3O_{7-x}$ | $BaCO_3$; $CuO$; $SrCO_3$; $Y_2O_3$ | 950 | 8 | Y-Ba-Cu-O-* | 10.1016/S0921-4534(00)01748-2 |
| $YBa_{2-x}Sr_xCu_3O_{7-x}$ | $BaCO_3$; $CaCO_3$; $CuO$; $Y_2O_3$ | 950 | 8 | Y-Ba-Cu-O-* | 10.1016/S0921-4534(00)01748-2 |
| $Y_{1-x}Sm_xBa_2Cu_3O_{7-x}$ | $Ba_2CO_3$; $CuO$; $Sm_2O_3$; $Y_2O_3$ | 940 | 90 | Y-Ba-Cu-O-* | 10.1016/s0025-5408(01)00539-6 |
| $Y_{1-x}Pr_xBa_2Cu_3O_y$ | $BaCO_3$; $CuO$; $Pr_2O_3$; $Y_2O_3$ | 930 | 96 | Y-Ba-Cu-O-* | 10.1016/j.physc.2008.05.031 |
| $(Y_{1-x}Ca_x)SrBaCu_{2.80}(PO_4)_{0.20}O_y$ | $BaO$; $CaCO_3$; $CuO$; $NH_4H_2PO_4$; $SrCO_3$; $Y_2O_3$ | 1000 | 32 | Y-Ba-Cu-O-* | 10.1016/S0921-4534(01)00104-6 |
| $Y_{1-x}Ca_xBa_{2-x}La_xCu_3O_y$ | $BaCO_3$; $CaCO_3$; $CuO$; $La_2O_3$; $Y_2O_3$ | 930 | 72 | Y-Ba-Cu-O-* | 10.1016/j.physc.2009.05.010 |
| $Y_{1-x}(Yb_{0.9}Nd_{0.1})_xBa_2Cu_3O_z$ | $BaO$; $CuO$; $Nd_2O_3$; $Y_2O_3$; $Yb_2O_3$ | 910 | 12 | Y-Ba-Cu-O-* | 10.1016/j.physc.2009.05.019 |
| $Y_{1-x}Pr_xBa_2Cu_3O_y$ | $BaCO_3$; $CuO$; $Pr_2O_3$; $Y_2O_3$ | 930 | 96 | Y-Ba-Cu-O-* | 10.1016/j.physc.2009.05.119 |
| $Y_2Ba_4CuNbO_y$ | $BaCO_3$; $CuO$; $Nb_2O_5$; $Y_2O_3$ | None | None | Y-Ba-Cu-O-* | 10.1016/j.physc.2009.05.194 |
| $Y_xNd_{1-x+y}Ba_{2-y}Cu_3O_{6+x}$ | $BaCO_3$; $CuO$; $Nd_2O_3$; $Y_2O_3$ | 1070 | 174 | Y-Ba-Cu-O-* | 10.1016/S0921-4534(01)00351-3 |
| $Y_{1-z}Ca_zBa_{2-z}La_zCu_3O_x$ | $BaCO_3$; $CaCO_3$; $CuO$; $La_2O_3$; $Y_2O_3$ | 1010 | 24 | Y-Ba-Cu-O-* | 10.1016/S0921-4534(01)00366-5 |
| $Y_{1-x}Ho_xBa_2Cu_3O_y$ | $(Y_{1-x}Ho_x)_2BaCuO_5$; $BaCuO_2$; $CuO$ | 550 | 40 | Y-Ba-Cu-O-* | 10.1016/S0921-4534(01)00368-9 |
| $YBa_2Cu_3F_{0.4}O_x$ | $YBa_2Cu_3F_4O_x$; $YBa_2Cu_3 O_x$ | 900 | 8 | Y-Ba-Cu-O-* | 10.1016/S0924-0136(99)00474-4 |
| $YBa_2Cu_{3-x}In_xO_y$ | $BaCO_3$; $CuO$; $In_2O_3$; $Y_2O_3$ | 1233 | 24 | Y-Ba-Cu-O-* | 10.1016/j.physc.2010.05.073 |
| $Y_{1-x}Pr_xBa_2Cu_3O_{7-d}$ | $BaCO_3$; $CuO$; $Pr_6O_{11}$; $Y_2O_3$ | 1213 | 183 | Y-Ba-Cu-O-* | 10.1021/cm9604928 |
| $YBa_2Co_xCu_{3-x}O_{7-x}$ | $BaCO_3$; $Co_2O_3$; $CuO$; $Y_2O_3$ | 980 | 25 | Y-Ba-Cu-O-* | 10.1016/j.catcom.2006.11.029 |
| $YBa_2Cu_{3-x} Co_xO_y$ | $BaCO_3$; $Co_2O_3$; $CuO$; $Y_2O_3$ | 900 | 48 | Y-Ba-Cu-O-* | 10.1016/S0921-4534(01)01286-2 |
| $YBa_2Cu_{3-x}M_xO_y$ | $BaCO_3$; $CuO$; $Fe_2O_3$; $Y_2O_3$ | 900 | 48 | Y-Ba-Cu-O-* | 10.1016/S0921-4534(02)01268-6 |

| Compound | Precursors | Temp (°C) | Time (h) | System | DOI |
|---|---|---|---|---|---|
| $YBa_2Cu_{3-x}M_xO_y$ | $BaCO_3$; $Co_3O_4$; $CuO$; $Y_2O_3$ | 900 | 48 | Y-Ba-Cu-O-* | 10.1016/S0921-4534(02)01268-6 |
| $YBa_{2-x}La_xCu_{3-x}Al_xO_z$ | $Al_2O_3$; $BaCO_3$; $CuO$; $La_2O_3$; $Y_2O_3$ | 940 | 66 | Y-Ba-Cu-O-* | 10.1016/j.physc.2012.01.013 |
| $Y_xGd_{1-x}Ba_2Cu_3O_{7-x}$ | $BaCO_3$; $CuO$; $Gd_2O_3$; $Y_2O_3$ | 950 | 58 | Y-Ba-Cu-O-* | 10.1016/S0921-4534(02)01441-7 |
| $YBa_{2-x}La_xCu_{3-x}Zn_xO_z$ | $BaCO_3$; $CuO$; $La_2O_3$; $Y_2O_3$; $ZnO$ | 940 | 66 | Y-Ba-Cu-O-* | 10.1016/j.physc.2012.01.013 |
| $Y_{1-x}Ca_xBa_2Cu(Cu_{1-y}Mg_y)_3O_{7-x}$ | $BaCO_3$; $CaCO_3$; $CuO$; $MgO$; $Y_2O_3$ | 940 | 24 | Y-Ba-Cu-O-* | 10.1016/j.physc.2012.02.031 |
| $Y_{1-x}Sm_xBa_2Cu_3O_{7-x}$ | $Ba_2CO_3$; $CuO$; $Sm_2O_3$; $Y_2O_3$ | 940 | 90 | Y-Ba-Cu-O-* | 10.1016/S0025-5408(01)00539-6 |
| $YBaCuFeO_5$ | $BaCO_3$; $CuO$; $Fe_2O_3$; $Y_2O_3$ | 1050 | 60 | Y-Ba-Cu-O-* | 10.1103/PhysRevB.91.064408 |
| $Y_{1-x}Pr_xBa_2Cu_3O_{7-x}$ | $BaCO_3$; $CuO$; $Pr_6O_{11}$; $Y_2O_3$ | 935 | 36 | Y-Ba-Cu-O-* | 10.1016/S0167-577X(01)00577-8 |
| $Y_2Ba(Cu_{1-x}Zn_x)O_5$ | $BaCO_3$; $CuO$; $Y_2O_3$; $ZnO$ | 950 | 30 | Y-Ba-Cu-O-* | 10.1016/s0955-2219(01)00097-8 |
| $YBa_2Cu_{3-x}Gd_xO_{7-x}$ | $BaCO_3$; $CuO$; $Gd_2O_3$; $Y_2O_3$ | 940 | 20 | Y-Ba-Cu-O-* | 10.1007/s10854-012-0917-0 |
| $Ba(Zr_{0.84}Y_{0.15}Cu_{0.01})O_{3-x}$ | $BaCO_3$; $CuO$; $Y_2O_3$; $ZrO_2$ | 1500 | 12 | Y-Ba-Cu-O-* | 10.1016/j.ijhydene.2014.02.072 |
| $Y_{2-x}Dy_xBaCuO_5$ | $BaCO_3$; $CuO$; $DY_2O_3$; $Y_2O_3$ | 1000 | 60 | Y-Ba-Cu-O-* | 10.1016/j.ssc.2004.02.026 |
| $YBa_2Cu_{3-x}Al_xO_{7-x}$ | $Al_2O_3$; $BaCO_3$; $CuO$; $Y_2O_3$ | 550 | 24 | Y-Ba-Cu-O-* | 10.1016/S0921-4534(02)02057-9 |
| $Y_{0.85}Ca_{0.15}Ba_2Cu_3O_{7-x}$ | $BaCO_3$; $CaCO_3$; $CuO$; $Y_2O_3$ | 930 | 24 | Y-Ba-Cu-O-* | 10.1016/j.ssc.2004.03.002 |
| $(Y_{1-x-y}Pr_xCa_y)Ba_2Cu_3O_{7-x}$ | $BaCO_3$; $CaCO_3$; $CuO$; $Pr_6O_{11}$; $Y_2O_3$ | 940 | 72 | Y-Ba-Cu-O-* | 10.1016/S0921-4534(02)02362-6 |
| $Y_2Ba(Cu_{1-x}Ni_x)O_5$ | $BaCO_3$; $CuO$; $NiO$; $Y_2O_3$ | 1300 | 32 | Y-Ba-Cu-O-* | 10.1016/s0921-5107(00)00566-3 |
| $YBa_2Cu_{3-x}Gd_xO_{7-x}$ | $BaCO_3$; $CuO$; $Gd_2O_3$; $Y_2O_3$ | 940 | 20 | Y-Ba-Cu-O-* | 10.1007/s10854-012-1022-0 |
| $Tl_2Ba_2Ca_{1-x}Y_x(Cu_{1-y}Co_y)_2O_8$ | $Ba_2Ca_{1-x}Y_x(Cu_{1-y}Co_y)O_{4+x}$; $CoO$; $CuO$; $Tl_2O_3$ | 930 | 30 | Y-Ba-Cu-O-* | 10.1016/S0921-4534(03)00628-2 |
| $Y_{1-x}Eu_xBa_2Cu_3O_{7-x}$ | $BaCO_3$; $CuO$; $Eu_2O_3$; $Y_2O_3$ | 1015 | 32 | Y-Ba-Cu-O-* | 10.1016/S0921-4534(03)00704-4 |
| $Y(Ba_{2-x}R_x)Cu_3O_{7-x}$ | $BaCO_3$; $CuO$; $La_2O_3$; $Y_2O_3$ | 940 | 72 | Y-Ba-Cu-O-* | 10.1016/S0921-4534(03)00810-4 |
| $Y(Ba_{2-x}R_x)Cu_3O_{7-x}$ | $BaCO_3$; $CuO$; $Nd_2O_3$; $Y_2O_3$ | 940 | 72 | Y-Ba-Cu-O-* | 10.1016/S0921-4534(03)00810-4 |

| Compound | Precursors | Temp (°C) | Time (h) | Class | DOI |
|---|---|---|---|---|---|
| Y(Ba$_{2-x}$R$_x$)Cu$_3$O$_{7-x}$ | BaCO$_3$; CuO; Pr$_6$O$_{11}$; Y$_2$O$_3$ | 940 | 72 | Y-Ba-Cu-O-* | 10.1016/S0921-4534(03)00810-4 |
| YBa$_{2-x}$Sm$_x$Cu$_3$O$_{7-x}$ | BaCO$_3$; CuO; Sm; Y$_2$O$_3$ | 935 | 24 | Y-Ba-Cu-O-* | 10.1016/j.physc.2018.02.026 |
| Y$_{1-y-x}$Co$_y$Ca$_x$Ba$_2$Cu$_3$O$_{7-x}$ | BaCO$_3$; CaCO$_3$; Co$_3$O$_4$; CuO; Y$_2$O$_3$ | None | None | Y-Ba-Cu-O-* | 10.1016/j.physc.2018.02.029 |
| Y$_{0.98-x}$Co$_{0.02}$Ca$_x$Ba$_2$Cu$_3$O$_{7-x}$ | BaCO$_3$; CaCO$_3$; Co$_3$O$_4$; CuO; Y$_2$O$_3$ | 950 | 24 | Y-Ba-Cu-O-* | 10.1016/j.physc.2018.02.029 |
| YBa$_2$Cu$_3$(OH)$_x$ | Ba(OC$_3$H$_7$)$_2$; Cu(CH$_3$COO)$_2$; Y(OC$_3$H$_7$)$_3$ | None | None | Y-Ba-Cu-O-* | 10.1111/j.1551-2916.2008.02900.x |
| YBa$_2$(Cu$_{1-x}$Cr$_x$)$_4$O$_8$ | Ba(CH$_3$COO)$_2$; Cr(NO$_3$)$_3$·9H$_2$O; Cu(CH$_3$COO)$_2$·H$_2$O; Y$_2$O$_3$ | 800 | 50 | Y-Ba-Cu-O-* | 10.1016/j.chemphys.2006.12.001 |
| Y$_3$Ba$_5$Ca$_2$Cu$_8$O$_{18}$ | Ba(NO$_3$)$_2$; CaCO$_3$; CuO; Y$_2$O$_3$ | 900 | 72 | Y-Ba-Cu-O-* | 10.1016/j.solidstatesciences.2011.08.024 |
| Y$_{1-x}$Ca$_x$BaSrCu$_3$O$_y$ | BaCO$_3$; CaCO$_3$; CuO; SrCO$_3$; Y$_2$O$_3$ | 1233 | 24 | Y-Ba-Cu-O-* | 10.1016/S0921-4534(03)01167-5 |
| BaCe$_{0.5}$Zr$_{0.3}$Y$_{0.08}$Yb$_{0.08}$Cu$_{0.04}$O$_{3-x}$ | BaCO$_3$; CeO$_2$; CuO; Y$_2$O$_3$; Yb$_2$O$_3$; ZrO$_2$ | 1400 | 3 | Y-Ba-Cu-O-* | 10.1016/j.ijhydene.2015.05.020 |
| Y$_{1-x}$Ca$_x$Ba$_{1.9}$Nd$_{0.1}$Cu$_3$O$_y$ | BaCO$_3$; CaO; CuO; Nd$_2$O$_3$; Y$_2$O$_3$ | 950 | 36 | Y-Ba-Cu-O-* | 10.1016/S0921-4534(03)01275-9 |
| Fe$_{0.5}$Cu$_{0.5}$Ba$_2$YCu$_2$O$_{7+x}$ | BaCO$_3$; CuO; Fe$_2$O$_3$; Y$_2$O$_3$ | 930 | 110 | Y-Ba-Cu-O-* | 10.1016/S0921-4534(03)01294-2 |
| YBa$_{2-x}$La$_x$Cu$_3$O$_y$ | BaCO$_3$; CuO; La$_2$O$_3$; Y$_2$O$_3$ | 920 | 42 | Y-Ba-Cu-O-* | 10.1016/s0038-1098(00)00360-4 |
| Fe$_x$Cu$_{1-x}$Ba$_2$YCu$_2$O$_{7+y}$ | BaCO$_3$; CuO; Fe$_2$O$_3$; Y$_2$O$_3$ | 930 | 110 | Y-Ba-Cu-O-* | 10.1016/j.ssc.2005.03.017 |
| (Y$_{2-x}$Sm$_x$)Ba(Cu$_{1-y}$Co$_y$)O$_5$ | BaCO$_3$; CoO; CuO; Sm$_2$O$_3$; Y$_2$O$_3$ | 850 | 32 | Y-Ba-Cu-O-* | 10.1016/s0955-2219(03)00179-1 |
| YBa$_2$Cu$_3$F$_{0.4}$O$_x$ | YBa$_2$Cu$_3$F$_4$O$_x$; YBa$_2$Cu$_3$O$_x$ | 900 | 8 | Y-Ba-Cu-O-* | 10.1016/s0924-0136(99)00474-4 |
| YBa$_2$(Cu$_{1-x}$Ni$_x$)$_3$O$_{7-x}$ | BaCO$_3$; CuO; Ni$_2$O$_3$; Y$_2$O$_3$ | None | None | Y-Ba-Cu-O-* | 10.1016/s0038-1098(01)00490-2 |